\documentclass[aps,pra,nobalancelastpage,superscriptaddress, 10pt, twocolumn]{revtex4-2}

\usepackage{color,ifthen,amsthm,amsmath,amsxtra,amsfonts,dsfont,graphicx,bm,tikz,scalerel,wasysym,bbm,graphicx,amsthm,braket,physics,empheq,
comment}
\usepackage[inline]{enumitem}
\usepackage[colorlinks=true,linkcolor=blue, citecolor=blue, urlcolor=blue, bookmarks]{hyperref}

\newcommand{\be}{\begin{equation}}
\newcommand{\ee}{\end{equation}}
\newcommand{\bea}{\begin{eqnarray}}
\newcommand{\eea}{\end{eqnarray}}
\newcommand{\bse}{\begin{subequations}}
\newcommand{\ese}{\end{subequations}}

\theoremstyle{plain}

\theoremstyle{plain}

\theoremstyle{plain}

\theoremstyle{plain}

\begin{document}

\title{Full Counting Statistics of  Charge in Quenched Quantum Gases}

\author{D\'avid X. Horv\'ath}
\affiliation{Department of Mathematics, King’s College London, Strand, London WC2R 2LS, United Kingdom}
\affiliation{SISSA and INFN Sezione di Trieste, via Bonomea 265, 34136 Trieste, Italy}
\author{Colin Rylands}
\affiliation{SISSA and INFN Sezione di Trieste, via Bonomea 265, 34136 Trieste, Italy}

\begin{abstract}
Unless constrained by symmetry,   measurement of an observable on an ensemble of identical quantum systems returns a distribution of values which are encoded in the full counting statistics.  While the mean value of this distribution is important for determining certain properties of a system,  the full distribution can also exhibit universal behavior.  In this paper we study the full counting statistics of particle number in one dimensional interacting Bose and Fermi gases which have been quenched far from equilibrium.  In particular we consider the time evolution of the Lieb-Liniger and Gaudin-Yang models quenched from a Bose-Einstein condensate initial state and calculate the full counting statistics of the particle number within a subsystem.   We show that the scaled cumulants of the charge in the initial state and at long times are simply related and in particular the latter are independent of the model parameters. Using the quasi-particle picture we obtain the full time evolution of the cumulants and find that although their endpoints are fixed, the finite time dynamics depends strongly on the model parameters. We go on to construct the scaled cumulant generating functions and from this determine the limiting charge probability distributions at long time which are shown to exhibit distinct, non-trivial and non-Gaussian fluctuations and large deviations.

\end{abstract}

\maketitle

\section{Introduction}
\label{sec:intro}
Symmetry and universality are two of the most powerful concepts in theoretical physics. The former can dramatically reduce the complexity of systems, places stringent constraints on allowed physical processes and gives rise to conservation laws via Noether's theorem. The latter instead explains why vastly different systems can display near identical features and how  simple physical principles can underpin many seemingly complex phenomena.  These concepts have been extensively studied in the context of closed quantum systems which are close to equilibrium leading to the discovery of many ubiquitous properties and the development of numerous powerful and widely applicable techniques of analysis.  In recent years however questions of universality and its emergence in far from equilibrium systems have come to the fore and in this context one-dimensional integrable models have been widely studied~\cite{PolkovnikovReview, calabrese2016introduction, VidmarRigol, essler2016quench, doyon2020lecture, bastianello2022introduction, alba2021generalized,rylands2020nonequilibrium}. These models possess special symmetry properties endowing them with an infinite number of mutually commuting conserved charges thereby placing strong constraints on their dynamics~\cite{korepin1997quantum}.  At the same time  these properties facilitate  exact analytic solutions of the models through Bethe ansatz techniques allowing in-depth analysis of their thermodynamic properties~\cite{takahashi1999thermodynamics}.  

Despite this analytic control however, uncovering the non-equilibrium properties of integrable models remains challenging and is still a highly active  area of research.  Nevertheless many universal features have been established, in particular concerning the dynamics within a subsystem where it has been shown that the system relaxes locally to a stationary state described by a generalized Gibbs ensemble (GGE)~\cite{PolkovnikovReview, calabrese2016introduction, VidmarRigol, essler2016quench, doyon2020lecture, bastianello2022introduction, alba2021generalized}.  Building upon this a number of exact techniques have been developed including the quench action method~\cite{caux2013time,caux2016quench} which allows one to determine this GGE explicitly and generalized hydrodynamics (GHD) which describes the long time  and large scale dynamics of an inhomogeneous state.~\cite{castroalvaredo2016emergent,bertini2016transport,Doyon_LongRangeCorrelations}.

A significant driver of interest in the non-equilibrium dynamics of integrable models has been the advent of numerous experimental platforms which allow for the simulation of isolated many-body quantum systems with a high degree of accuracy and control~\cite{bloch2008many, cazalilla2011one, guan2013fermi}.  Chief amongst these are ultra cold atomic gas setups which have the ability to faithfully simulate integrable systems.  The Lieb-Liniger model~\cite{lieb1972finite} of interacting bosons,  the Gaudin-Yang model~\cite{yang1967some,gaudin1967systeme} of interacting fermions and the sine-Gordon field theory~\cite{zamolodchikov1979factorized} are all well known integrable models which can be simulated within cold atom experiments~\cite{kinoshita2006quantum,gritsev2007spectroscopy,gritsev2007linear,cazalilla2011one,guan2013fermi,pigneur2018relaxation,zache2020extracting,roy2019quantum, Horvath_2019,roy2021quantum,rylands2020photon,Horvath_2022,bastianello2023sinegordon,senaratne2022spin}.   The relaxation of such models to GGEs~\cite{langen2015experimental} as well as the vailidty of GHD~\cite{schemmer2019generalized,malvania2021generalized,le2023observation} has been observed and tested extensively in such experiments ensuring that despite their apparent fine tuned nature integrable models are the appropriate description of these systems. 

\begin{figure}[t]
\centering
\includegraphics[trim= 0 100 0 0, width=\columnwidth]{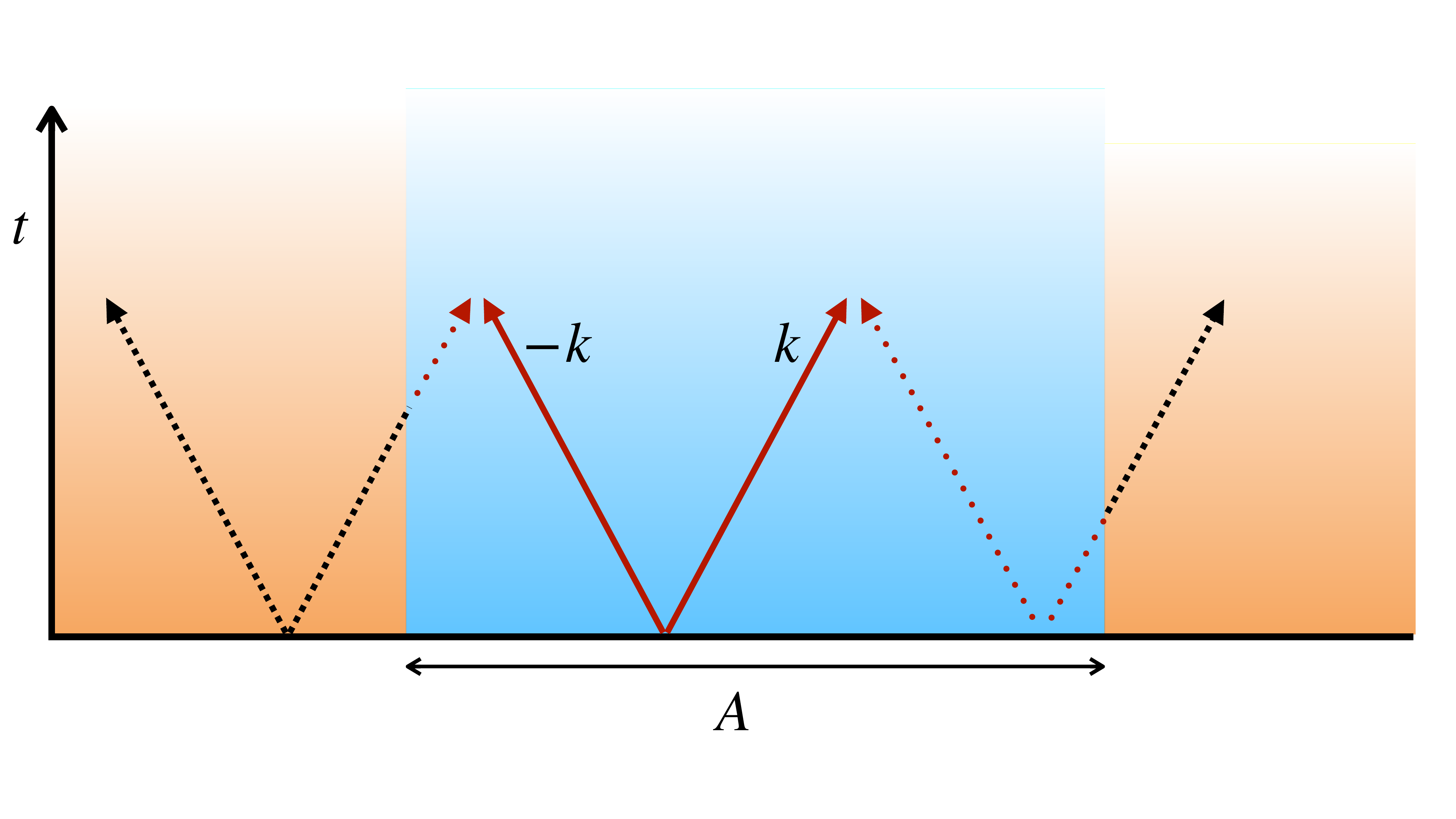}
\caption{\label{fig:QPpicture}
The quasi-particle picture of charge fluctuations within a region $A$.  The quench causes pairs of correlated  quasi-particles with equal and opposite momenta, $k,-k$ to be emitted from the initial state.  The pairs propagate throughout the system, carrying with them the charge. A quasi-particle pair’s contribution to charge fluctuation differs depending on whether both members of the pair (solid lines) or a single member (dotted lines) are inside $A$.  At short times the fluctuations inside $A$ are given by pairs while times which are long compare to the subsystem size only a single member of a pair can contribute.}
\end{figure}

In this work we shall study the interplay between non-equilibrium dynamics of integrable models and the fluctuations of their charge within a subsystem in the context of cold atom gas experiments.   We do this by calculating the full counting statistics (FCS) of the particle number in the Lieb-Liniger (LL) and Gaudin-Yang (GY) models. The systems shall be taken out of equilibrium by quenching them from an initial Bose-Einstein condensate (BEC) state which is not only experimentally relevant and conceptually simple but also allows us to characterize the dynamics exactly.  We shall show that there exists a simple relationship between the cumulants of charge in the initial state and the GGE to which these systems relax at long times.  This relationship is in fact universal, relying only on the presence of integrability and not on details of the model.  However,  while the cumulants in the final state are fixed by the initial state their decomposition in terms of quasi-particle excitations differs from model to model meaning that the finite time dynamics will differ.  Specifically, we find that the connected $m$-point function of the charge,  $N$ in a subsystem, $A$  at the initial and final times are related by a factor of $1/2^{m-1}$,  that is 
\begin{eqnarray}
\lim_{|A|\to\infty}\lim_{t\to\infty}\frac{\left<\right.\!N^m_A(t)\!\left.\right>^c}{\left<\right.\!N^m_A(0)\!\left.\right>^c}=\frac{1}{2^{m-1}}.
\end{eqnarray}
We provide a general derivation of this result below along with an explanation in terms of the qusaiparticle picture~\cite{calabrese2005evolution,alba2017entanglement}(see Fig. \ref{fig:QPpicture}) and explicit checks in our models of choice. Moreover, since the leading order (linear in subsystem size) behavior of the cumulants of charge fluctuations in the GGEs are the same in all cases they define the same limiting coarse-grained or continuous probability distributions (PDs) which can be explicitly determined. Nevertheless, the microscopic PDs (retaining information about sub-leading corrections) can be different. This fact arises from the transmutation of particle statistics in the different parameter regimes of each model and manifests in a simple and intuitive way. For example, in the strongly attractive Fermi gas the fermions are bound into tight pairs forming hard core bosons with double the charge, a fact which is reflected in vanishing probabilities for observing an odd number of particles in a subsystem. 

The rest of the paper is arranged as follows: In Sec.~\ref{sec:noneq} we review our object of study, the full counting statistics and some basic properties of cumulant generating functions and their associated probability distributions. We also briefly recall some properties of integrable models and prove a general relationship between the FCS calculated in a GGE and in the diagonal ensemble. In Sec.~\ref{sec:models} we introduce the models we shall study as well as the particular initial states of interest. In Sec.~\ref{sec:FCSLL} we determine the cumulants of charge in Lieb-Liniger model for both repulsive and attractive regimes. In Sec.~\ref{sec:FCSGY} we carry out an analogous calculation  in  the repulsive and attractive regimes of the Gaudin-Yang model. In the penultimate section we study the analytic properties of the scaled cumulant generating functions for both our models. We use this knowledge to determine the limiting charge probability distributions in both the initial and final states. In the last section we summarize our work and discuss open questions and potential future directions. 

\section{Non-equilibrium Full Counting Statistics in integrable models}
\label{sec:noneq}

\subsection{Full counting statistics and charge probability distributions}

We wish to characterize the fluctuations of the particle number in the LL and GY models which have been quenched from BEC states.  While these charges are conserved quantities for the full system,  when we restrict to a subsystem they exhibit nontrivial fluctuations and dynamics~\cite{cherng2007quantum,klich2009quantum,eisler2013full,eisler2013universality,lovas2017full,najafi2017full,collura2017full,bastianello2018from,arzamasovs2019full,vannieuwkerk2019self,perfetto2020dynamics,calabrese2020full,collura2020how,doyon2022ballistic,oshima2022disordered,tartaglia2022real,parez2021quasiparticle,parez2021exact,bertini2022nonequilibrium,bertini2023dynamics,ScopaDXH_2022}.  To study this we calculate the full counting statistics of the charge  within a region $A$ of length $\ell$. We take $\ell$ to be large but much smaller than the full system, $1\ll\ell\ll L$ where $L$ is the full system size and in the end take the thermodynamic limit $\ell,L\to \infty$. For a charge operator $\hat{N}$ whose restriction to the subsystem is denoted $\hat{N}_A$ the FCS are defined as 
\begin{eqnarray}
Z_\beta(A,t)=\tr[\rho_A(t)e^{\beta \hat{N}_A}]
\end{eqnarray}
where $\rho_A(t)=\tr_{\bar{A}}[\rho(t)]$ is the reduced density matrix of $A$ at time $t$ and $\rho(t)=\ketbra{\Psi_0(t)}{\Psi_0(t)}$ is the density matrix of the full system which has been quenched from the intial state $\ket{\Psi_0}\to\ket{\Psi_0(t)}=e^{-i Ht}\ket{\Psi_0}$.  
 While the FCS of charge and related quantities like the work distribution have been studied previously in certain interacting integrable models~\cite{palmai2014quench,groha2018full,calabrese2020full,rylands2019quantum,rylands2019loschmidt,perfetto2019quench,bastianello2018exact}, recent developments have allowed to study these quantities in far from equilibrium systems as well as their associated current statistics~\cite{Doyon_2019,myers2020transport,doyon2022ballistic, gopalakrishnan2022theory,krajnik2022exact,krajnik2022universal, samajdar2023quantum, krajnik2023universal, mcculloch2023full,rylands2023transport}.

From the FCS we can determine many properties of the dynamics of the particle number fluctuations within the subsystem and its interplay with the relaxation of the system to its long time steady state. 
Specifically, we can calculate the connected correlation functions of the charge via the cumulants of the FCS,
\begin{eqnarray}
\left<N_A^m(t)\right>^c=\partial_\beta^m\log Z_\beta(A,t)|_{\beta=0}
\end{eqnarray}
for $m\in \mathbb{N}$ and where $\left<\cdot\right>^c$ means the connected part of the correlation function e.g. $\left<N_A^2(t)\right>^c=\left<N_A^2(t)\right>-\left<N_A(t)\right>^2$.  Moreover, by continuing $\beta \to i\beta$ we can obtain the charge probability distribution $P(n,t)$, 
\begin{eqnarray}
P(n,t)=\int_{-\pi}^\pi\frac{{\rm d}\beta}{2\pi}e^{-i\beta n}Z_{i\beta}(A,t)
\label{ProbFromLogZ}
\end{eqnarray}
which gives the  probability that a measurement of $\hat{N}_A$ at time $t$ returns the value $n$.  The latter is a natural quantity to study from the experimental point of view and particularly so in the case of cold atom experiments.  While the expectation value of an observable such as an order parameter is a central quantity to understand the nature of a system,  the full distribution of measurement outcomes is also enlightening and can unveil universal behavior. In cold atom experiments large numbers of measurements are required to be performed and so the full probability distribution is naturally obtained~\cite{armijo2010probing,jacmin2011sub,mazurenko2017cold,herce2023full}.  

By our analytical methods which we shall introduce shortly, the computation of the cumulants or the cumulant generating function  $\log Z_\beta(A,t)$ is only feasible 
if the length of the subsystem $\ell$ is also infinitely large.  In particular, the moment generating function for an extensive quantity in a large subsystem can be generally written as
\begin{equation}
\lim_{\ell\rightarrow \infty} Z_\beta(A,t)= e^{\ell C^s(\beta,t)+o(\ell)}
\end{equation}
where $C^s(\beta,t)$ is the scaled cumulant generating function (SCGF) defined as
\begin{equation}
\lim_{\ell\rightarrow \infty}\ell^{-1}\log Z_\beta(A,t)= C^s(\beta,t)
\end{equation}
and whose derivatives w.r.t. $\beta$ are the scaled cumulants
\begin{equation}
\partial^m_{\lambda} C^s(\beta,t)|_{\lambda=0}=\kappa^s_m(t)
\end{equation}
denoted by $\kappa^s_m$.
Therefore it is the scaled cumulants and their generating function, i.e., quantities with extensive scaling w.r.t. the subsystem size, which can be computed. From the SCGF, we can obtain the large $\ell$ limiting probability distribution (PD) for the charge fluctuations via the rate function $I(z)$ through 
\begin{equation}
P(\ell z,\ell t)\asymp e^{-\ell I(z,t)}
\label{LimitingPD}
\end{equation}
where $\asymp$ is understood as 
\begin{equation}
a(\ell,x)\asymp b(\ell,x),\,\,\text{ if } \,\, \lim_{\ell \rightarrow \infty} \frac{\log\, a}{\log\, b}=1\,,
\label{AsymptoticEquality}
\end{equation} and we assumed the same scaling for the time variable $t$.
Importantly the rate function $I(z)$ can be obtained as the Legendre-Fenchel transform of the SCGF according to the G\"artner-Ellis theorem.
The rate function also governs the large deviations of the limiting coarse-grained  probability distribution, which is a continuous distribution even if the microscopic PD is discrete. 
Whereas in a strict sense our methods allow for the computation of $I(z)$ via the SCGFs, we shall also use Eq. \eqref{ProbFromLogZ} and approximate $Z_{i\beta}(A,t=0,\infty$) at the initial and the steady-states by $\exp\left(\ell C^s(i\beta, t=0,\infty)\right)$. Doing so retains the discrete nature of charge fluctuations and allows for the onset of some interesting microscopic effects, which vanish in the $\ell \rightarrow \infty$ limit but are expected to be observed in large but finite subsystems.

\subsection{Review of integrable models and the Thermodynamic Bethe Ansatz}

In the subsequent sections we shall explicitly calculate the scaled cumulants $\kappa_m^s(t)$ as well as SCGF $C^s(\beta,t=0,\infty)$ in the LL and GY models however before this we shall derive some generic properties which will allow us to relate the scaled cumulants in the initial state, denoted by 
$\kappa^s_m(0)$, 
and in the GGE, denoted by $\kappa^s_m(\infty)$.  
To do this we keep things general and recall briefly some basic properties of integable models: Integrable models posses an extensive number of conserved quantities charges whose associated operators $\hat{Q}^{(k)},~k=1,2,\dots$ commute with the Hamiltonian. These imbue the model with a stable set of quasi-particle excitations which are indexed by a  discrete species index $n=1,\dots,N_s$ and parameterized by a continuous rapidity $\lambda^{(n)}_j$, $j=1,\dots,M_{n}$.  An eigenstate of the model is specified by the rapidities of the quasi-particles which are present, and we denote it by
\begin{equation}
|\bm{\lambda}\rangle= |\lambda^{(1)}_1,\ldots,\lambda^{(1)}_{M_1}; \ldots; \lambda^{(N_s)}_1,\ldots,\lambda^{(N_s)}_{M_{N_s}}\rangle\,.
\label{eq:basis}
\end{equation}
These states are also simultaneous eigenstates of all the conserved charges of the model, namely, 
\be
\hat{Q}^{(k)} \ket{\bm{\lambda}} = \sum_{n=1}^{N_s}\sum_{j=1}^{M_m} q_n^{(k)}(\lambda^{(m)}_j) \ket{\bm{\lambda}}.
\ee 
The first three conserved charges are chosen to coincide with the particle number $\hat{N}=\hat{Q}^{(0)}$, momentum $\hat{P}=\hat{Q}^{(1)}$ and Hamiltonian $\hat{H}=\hat{Q}^{(2)}$.  For simplicity we shall denote the charge, momentum and energy of a quasi-particle of species $n$ and rapidity $\lambda$ by $q_n=q^{(0)}_n(\lambda),p_n(\lambda)=q^{(1)}_n(\lambda)$ and $\epsilon_n(\lambda)=q^{(2)}_n(\lambda)$.   
In the thermodynamic limit and at finite density the model can be treated using the methods of the thermodynamics Bethe Ansatz (TBA)~\cite{takahashi1999thermodynamics}. Within this approach,  the quasi-particle content of a stationary state of the model can be encoded in the distributions $\rho_n(\lambda),\rho_n^h(\lambda),\rho_n^t(\lambda)$ and $\vartheta_n(\lambda)$ which are respectively the distribution of occupied quasiparticles of species $n$, the distribution unoccupied quasiparticles,  the total density of states $\rho_n^t(\lambda)=\rho_n(\lambda)+\rho_n^h(\lambda)$ and the occupation function $\vartheta_n(\lambda)=\rho_n(\lambda)/\rho^t_n(\lambda)$.  All of these functions are related to each other by the Bethe-Takahashi equations which take the form
\begin{eqnarray}
\rho^t_n(\lambda)=\frac{p_n'(\lambda)}{2\pi}-\sum_{l}\mathcal{T}_{nl}*\rho_l(\lambda)
\end{eqnarray}
where $(\cdot)'$ denotes differentiation w.r.t. $\lambda$ and $*$ is the convolution $f*g(y)=\int {\rm d}y f(x-y)g(y)$.  Here  $\mathcal{T}_{nl}(\lambda,
\mu)$ is the scattering kernel which characterizes the scattering between quasi-particles of type $n$ and $l$ with rapidities $\lambda$ and $\mu$.  In the cases of interest this is symmetric in the species index $\mathcal{T}_{nl}(x)=\mathcal{T}_{ln}(x)$.  

When the state contains many excitations the bare quasi-particle properties become dressed due to the interactions. These dressed quantities are denoted by $(\cdot)^{\rm dr}$ and satisfy the  integral equations
\begin{eqnarray}
    f_n^{\rm dr}(\lambda)=f_n(\lambda)-\sum_{l}\mathcal{T}_{nl}*[\vartheta_l(\lambda)f_l^{\rm dr}(\lambda)].
\end{eqnarray} 
Explicitly, the dressed charge is obtained from the solution of
\begin{eqnarray}
q^{\rm dr}_n(\lambda)=q_n-\sum_{l}\mathcal{T}_{nl}*[\vartheta_l(\lambda)q_l^{\rm dr}(\lambda)].
\end{eqnarray}
The quasi-particle velocity on the other hand is given by the ratio of dressed quantities, $v_n(\lambda)={\epsilon'}^{\rm dr}(\lambda)/{p'}^{\rm dr}(\lambda)$.  This can be recast into the equation
\begin{eqnarray}\label{eq:Vel}
\rho^t_n(\lambda)v_n(\lambda)=\frac{\epsilon_n'(\lambda)}{2\pi}-\sum_{l}\mathcal{T}_{nl}*[\rho_l(\lambda)v_l(\lambda)]
\end{eqnarray}
where we have used $\rho^t(\lambda)={p'}^{\rm dr}(\lambda)$. These sets of integral equations typically need to be solved numerically but in certain cases can be solved analytically as is the case for the LL model discussed below~\cite{denardis2014solution,piroli2016mulitparticle,piroli2016quantum}.

\subsection{Full counting statistics using TBA}

Having reviewed these basic properties and set up our notation we can now move on to determine the relationship between the FCS at $t=0$ and in the GGE.  We begin with the latter and define
\begin{align}\label{eq:fDef}
\lim_{t\to\infty}\tr[e^{\beta \hat{N}_A}\rho_{A}(t)]=\tr[e^{\beta \hat{N}_A}\rho_{\rm GGE}]\equiv\exp\left(\ell C^s_\text{GGE}(\beta)\right)
\end{align}
where we have neglected the $o(\ell)$ contributions and have introduced the generalized Gibbs ensemble
\begin{eqnarray}
\rho_{\rm GGE}=\frac{1}{\tr[e^{-\sum_k \beta^{(k)} \hat{Q}^{(k)}}]}e^{-\sum_k \beta^{(k)} \hat{Q}^{(k)}}.
\end{eqnarray}
Herein it is necessary to include both local and semi-local conserved charges in the GGE in order to obtain the correct description of the long time state~\cite{pozsgay2014correlations,wouters2014quenching,goldstein2014failure,ilievski2015complete,ilievski2016quasilocal}. The Lagrange multipliers above, $\beta^{(k)}$, are  determined by matching the expectation values of $\hat{Q}^{(k)}$ in $\rho_{\rm GGE}$ to those in the initial state.  To compute
$C^s_\text{GGE}(\beta)$ with $\beta\in \mathbb{R}$ we utilize the result of Ref. \cite{Doyon_2019} (see also~\cite{piroli2022thermodynamic,bertini2022nonequilibrium}) claiming that
\begin{equation}
\label{DoyonMeyersFCSAsFreeEDifference}
C^s_\text{GGE}(\beta)=f_\text{GGE}(\underline{\beta}^{(k)}-\beta)-f_\text{GGE}(\underline{\beta}^{(k)})
\end{equation}
where $f_\text{GGE}$ denotes the free energy density of a GGE characterized by the chemical potentials $\underline{\beta}^{(k)}$ and $\beta$ shifts one of the chemical potentials that corresponds to the conserved quantity under investigation. We can rewrite the above equations
in the typical manner of the thermodynamic Bethe ansatz 
by exchanging the trace 
for a path integral over the distributions $\rho_n(\lambda),\rho^h_n(\lambda)$~\cite{takahashi1999thermodynamics},
\begin{align}
e^{\ell C^s_\text{GGE}(\beta)}=\frac{\int \mathcal{D}[\rho(\lambda)]e^{-\ell \sum_n\int{\rm d}\lambda( g_n(\lambda)-\beta q_n)\rho_n(\lambda)-S_{YY}}}{\int \mathcal{D}[\rho(\lambda)]e^{-\ell \sum_n\int{\rm d}\lambda g_n(\lambda)\rho_n(\lambda)-S_{YY}}}
\label{EqSCFSSPathINtegral}
\end{align}
where $g_n(\lambda)=\sum_{k}\beta^{(k)}q^{(k)}_n(\lambda)$ and $S_{YY}$ is the Yang-Yang entropy which is proportional to $\ell$ as well and which counts how many microstates $\ket{\boldsymbol{\lambda}}$ correspond to the same macrostate given by the distributions $\rho_n(\lambda)$.  Evaluating this functional integral via a saddle point approximation we obtain
\begin{align}\label{eq:GGEsaddle}
C^s_\text{GGE}(\beta)=\!\sum_{n=1}^{N_s}\int \!\!{\rm d}\lambda\rho^t_n \!\!\left[\!\log\!\left(\frac{1+\eta^{-1}_{n,\beta}}{{1+\eta^{-1}_{n,0}}}\right)\!+\vartheta_n \!\log\frac{\eta_{n,\beta}e^{\beta q_n}}{\eta_{n,0}}\!\right]
\end{align}
with $\eta_{n,\beta}$ being given by
\begin{align}\label{eq:genericTBA}
\log \eta_{n,\beta}=-\beta q_n +g_n(\lambda)+\sum_{l}  \mathcal{T}_{nl}*\log[1+\eta^{-1}_{l,\beta}(\lambda)]\,,
\end{align}
while $\eta_{n,0}$ is evaluated in the same way at $\beta=0$ and can be related to the occupation function of the GGE via $\vartheta_n(\lambda)=1/(1+\eta_{n,0}(\lambda))$.  In order to evaluate $C^s_\text{GGE}(\beta)$ explicitly one must know the functional form of $g_n(\lambda)$ which can be done by making use of the quench action method.  Using this one can show that it is obtained from the extensive part the squared overlap between an eigenstate $\ket{\boldsymbol{\lambda}}$ and $\ket{\Psi_0}$ in the thermodynamic limit~\cite{alba2017renyi}. Namely,
\begin{eqnarray}\label{eq:overlapgeneral}
\lim_{\rm th}\frac{1}{L}\ln |\langle\boldsymbol{\lambda}|\Psi_0\rangle|^2=-\frac{1}{2}\sum_{n=1}^{N_s}\int {\rm d}\lambda\, g_n(\lambda)\rho_n(\lambda)\,. 
\end{eqnarray}
For the states that we consider the left hand side is known explicitly and the function $g_n(\lambda)$ is simply extracted from this.  

We now turn to the calculation of the FCS in the initial state.  In this instance, as discussed further in Appendix \ref{AppA}, the initial value can be obtained by considering the diagonal ensemble.  After expressing this as a path integral we find 
\begin{align}
e^{\ell C^s_0(\beta)}=\frac{\int \mathcal{D}[\rho(\lambda)]e^{-\frac{\ell}{2} \sum_n\int{\rm d}\lambda( g_n(\lambda)-2\beta q_n)\rho_n(\lambda)-\frac{1}{2}S_{YY}}}{\int \mathcal{D}[\rho(\lambda)]e^{-\frac{\ell}{2} \sum_n\int{\rm d}\lambda g_n(\lambda)\rho_n(\lambda)-\frac{1}{2}S_{YY}}}
\label{EqSCFISPathINtegral}
\end{align}
where the factor of $\frac{1}{2}$ in front of $S_{YY}$ arises from the fact that the  overlap with our initial state is non-zero only for parity invariant eigenstates which have half the entropy contribution.  Here we have denoted $C^s(\beta,0)$ by $C^s_0(\beta)$, which we express more explicitly as
\begin{align}\label{eq:DEsaddle}
C^s_0(\beta)=\frac{1}{2}\sum_{n=1}^{N_s}\int \!\!{\rm d}\lambda\bar{\rho}^t_n \!\!\left[\!\log\!\left(\frac{1+\bar{\eta}^{-1}_{n,\beta}}{{1+\bar{\eta}^{-1}_{n,0}}}\right)\!+\bar{\vartheta}_n \!\log\frac{\bar{\eta}_{n,\beta}e^{2\beta q_n}}{\bar{\eta}_{n,0}}\!\right]
\end{align}
with $\bar{\eta}_{n,\beta}$ being given by
\begin{align}\label{eq:genericTBA}
\log \bar{\eta}_{n,\beta}=-2\beta q_n +g_n(\lambda)+\sum_{l}  \mathcal{T}_{nl}*\log[1+\bar{\eta}^{-1}_{l,\beta}(\lambda)]\,.
\end{align}

Either comparing Eqs. \eqref{EqSCFSSPathINtegral} and \eqref{EqSCFISPathINtegral} or the TBA systems \eqref{eq:GGEsaddle} and \eqref{eq:DEsaddle}, we then find that 
\begin{eqnarray}
\label{eq:diagonalsaddle}
C^s_0(\beta)=\frac{1}{2}C^s_\text{GGE}(2\beta)\,,
\end{eqnarray}
where 
$C^s_\text{GGE}$ is given  by~\eqref{eq:GGEsaddle}. Therefore comparing with~\eqref{eq:fDef} we also find that the cumulants are related as
\begin{eqnarray}\label{eq:cumulantresult}
\kappa^s_m(\infty)=\frac{1}{2^{m-1}} \kappa^s_m(0). 
\end{eqnarray}
The above relations are one of the main results of this paper. We note that while~\eqref{eq:cumulantresult}
is a technical result whose proof relies upon the details of the TBA and quench action formalism, it also admits a very intuitive interpretation based upon the quasi-particle picture of entanglement dynamics~\cite{calabrese2005evolution, alba2017entanglement}, characterized in Fig.~\ref{fig:QPpicture}, which proceeds as follows. The initial state of the system can be expressed as a collection of correlated pairs of quasiparticles, of opposite momenta,   which are excited by the quench.   A pair of quasi-particles can then be viewed semi-classically as emerging from a single point in space. Its constituents propagate ballistically in opposite directions throughout the system thereby spreading correlations over a wider region.  The fluctuations of charge within the subsystem at $t=0$ are therefore governed by the distribution of pairs of quasiparticles which necessarily carry charge $2$ and are indexed by the absolute value of the momenta.  At long time, the quasiparticles have spread throughout the system and it is not possible for both members of an initially correlated pair to be inside the subsystem, thus the fluctuations of charge in the subsystem are governed by single quasi-particles of charge $1$ and indexed by the momentum rather than its absolute value.  Nevertheless, since the quasiparticle occupation function is a constant of the dynamics the distributions of pairs and unpaired quasi-particles are the same.  Thus we arrive at the expression~\eqref{eq:diagonalsaddle}  where we can recognize the $2\beta$ as originating from the charge of a pair of quasi-particles while the factor of $\frac{1}{2}$ comes from the fact that one sums over only the absolute value of the momenta.

A notable consequence of the relations is that since
the right hand sides of \eqref{eq:diagonalsaddle} and \eqref{eq:cumulantresult} are 
a property solely of the initial state then the left hand side is independent of the model specifics.  In the examples below we shall argue and then show explicitly that for BEC initial states 
$\kappa_m^s(0)=d$ $\forall m$ where $d$ is the average density
i.e.,  the charge is Poisson distributed and accordingly 
\begin{eqnarray}\label{eq:cumurelation}
\kappa_m^s(\infty)=\frac{d}{2^{m-1}}\,, 
\end{eqnarray}
which is exactly the same expression as the one obtained from the steady-state GGE  using \eqref{DoyonMeyersFCSAsFreeEDifference} as we shall demonstrate shortly.

Evidently~\eqref{eq:cumulantresult} does not tell us how the charge distribution evolves between its initial value and the GGE value which necessarily depends on the model.  However we can also characterize this in integrable models using recently obtained results on the charged moments and full counting statistics using the method of space-time duality~\cite{bertini2022nonequilibrium,bertini2023dynamics}. In particular, as discussed above the scaled cumulants obey a quasi-particle picture~\cite{calabrese2005evolution,alba2017entanglement} meaning that charge fluctuations are spread throughout the system via the ballistic propagation of pairs of quasi-particles which are shared between $A$ and $\bar{A}$ resulting in
\begin{eqnarray}\nonumber
\left<N^m_A(t)\right>^c &=& \sum_{n}\int \mathrm{d}\lambda\,{\rm min}[2 t |v_n(\lambda)|,\ell] \mathcal{K}^m_{n,\infty}(\lambda)\\\label{eq:timedepcum}
&&+\left(\ell-{\rm min}[2 t |v_n(\lambda)|,\ell]\right)\mathcal{K}^m_{n,0}(\lambda)
\end{eqnarray}
where we have introduced 
$\mathcal{K}^m_{n,{\rm x}}(\lambda)$
with ${\rm x}=0,\infty$ being the rapidity and species resolved scaled cumulant in the initial state or GGE, i.e.
\begin{eqnarray}
\kappa^s_m({\rm x})=\sum_n\int \mathrm{d}\lambda \,\mathcal{K}^m_{n,{\rm{x}}}(\lambda).
\end{eqnarray} 
Additionally, ${\rm min}[2 t |v_n(\lambda)|,\ell]$ is the characteristic function which counts the number of quasi-particle pairs shared between $A$ and $\bar{A}$,  while $\ell -{\rm min}[2 t |v_n(\lambda)|,\ell]$ can be interpreted as the number quasiparticle from pairs which are contained solely within the subsystem.  Thus, within the subsystem the charge fluctuations are governed either by pairs of quasi-particles both of whom are inside $A$ and whose contribution is given by $\mathcal{K}^m_{n,0}(\lambda)$  or single quasi-particles which originated from outside $A$ or whose partner has exited $A$ and which contribute $\mathcal{K}^m_{n,\infty}(\lambda)$.

The first cumulant returns the expectation value of the charge and is conserved which can be seen using~\eqref{eq:cumurelation} and~\eqref{eq:timedepcum} at $m=1$. The second cumulant gives the charge susceptibility which has a compact expression
\begin{eqnarray}
\mathcal{K}_{n,\infty}^{2}(\lambda)=[q_n^{\rm dr}(\lambda)]^2\rho_n(\lambda)(1-\vartheta_n(\lambda)).
\end{eqnarray}
Higher cumulants become more complicated and involve dressing of lower cumulants.  For example, the third cumulant is given by
\begin{eqnarray}\nonumber
\frac{\mathcal{K}_{n,\infty}^{3}(\lambda)}{\mathcal{K}_{n,\infty}^{2}(\lambda)}=q_n^{\rm dr}(\lambda)(1-2\vartheta_n(\lambda))-\frac{3 [\sum_{l}\mathcal{T}_{nl}*\mathcal{K}_{l,\infty}^{2}]^{\rm dr}(\lambda)}{q_n^{\rm dr}(\lambda)}
\end{eqnarray} 
where we see that it depends also on the dressed second cumulant.  We omit higher expressions for cumulants which can nevertheless be systematically computed. 

Before moving on, we make two brief remarks on the preceding discussion.  
First, the result~\eqref{eq:cumulantresult} relies upon the validity of the diagonal ensemble for calculating the initial state value which is discussed further in Appendix~\ref{AppA}. To check the validity of \eqref{eq:cumulantresult} we shall compute the scaled cumulants in the initial states by independent means and show that the results of the diagonal ensemble, involving 
 interaction-dependent functions, yield the same result. This is not always the case however and  the diagonal ensemble cannot be applied to initial states such that the subsystem is an eigenstate of the charge and hence has no fluctuations.  This is the case for the magnetization of the GY model quenched from the BEC state.   However for the models and initial states which we consider the diagonal ensemble correctly reproduces the initial FCS of the particle number. 
Second, although some analytic results exist for the distributions functions which determine \eqref{eq:GGEsaddle} and \eqref{EqSCFISPathINtegral}, their explicit evaluation is feasible only in certain  special limits of our models. In spite of this, at generic interactions, it is straightforward to numerically compute the (first few) scaled cumulants and check the validity of \eqref{eq:cumulantresult}. The numerical computation of the SCGF is also possible but instead of comparing the numerically obtained SCGFs against \eqref{eq:diagonalsaddle} on a strictly finite interval, we shall rather construct the SCGFs at arbitrary interactions as a Taylor series defined by the cumulants, which is justified by the exponentially decaying behavior of the scaled cumulants.

\section{Models and Setup}
\label{sec:models}
\subsection{Lieb-Liniger model}
\label{subsec:LL}
The standard model for describing one-dimensional interacting bosons in the context of cold atom experiments is the Lieb-Liniger model. The Hamiltonian is given by
\begin{equation}\label{eq:LLHam}
H_{LL}\!=\!\!\int_0^L \!\!\!\!{\rm d}x\, b^\dag(x)\!\!\left[-\frac{\partial_x^2}{2m}\right]\!b(x)+c b^\dag(x)b(x)b^\dag(x)b(x).
\end{equation}
Here $b^\dag(x),b(x)$ are canonical bosonic operators satisfying $[b(x),b^\dag(y)]=\delta(x-y)$ they describe bosons of mass $m$ which interact via a density-density interaction of strength $c$ on a system of length $L$.  We shall consider both the repulsive $c>0$ and attractive $c<0$ cases. From here on we take $m=1/2$ for simplicity and  assume periodic boundary conditions.  The model has a single $U(1)$ charge, the particle number $\hat{N}=\int {\rm d}x\, b^\dag(x)b(x)$.  

 In the repulsive  case there is only one species of quasi-particle  with bare charge, momentum and energy given by
\begin{eqnarray}
q=1,~p(\lambda)=\lambda,~\epsilon(\lambda)=\lambda^2.
\end{eqnarray}
As outlined above, the properties of a stationary state of $H$ are encoded in the distributions $\rho(\lambda),\rho^h(\lambda),\rho^t(\lambda)$ and $\vartheta(\lambda)$ which are connected via the Bethe equations
\begin{eqnarray}\label{eq:LLBAE}
\rho^t(\lambda)=\frac{1}{2\pi}+\int_{-\infty} ^\infty{\rm d} \mu\,a_2(\lambda-\mu)\rho(\mu)
\end{eqnarray}
where the scattering kernel is given by $a_n(x)= \frac{1}{2\pi}\frac{n|c|}{(nc/2)^2+x^2}$.  

In the attractive case the spectrum of the model is entirely different and there are an infinite number of quasi-particle species corresponding to bound states of $n$ bosons.  These quasi-particles have the following bare charge,  momentum and energy
\begin{eqnarray}
q_n=n,~p_n(\lambda)=n\lambda,
~\epsilon_n(\lambda)=n\lambda^2+\frac{|c|}{12}n(n^2-1). 
\end{eqnarray} 
We denote the distributions of the $n$-boson bound states by $\rho_n(\lambda),\rho_n^h(\lambda),\rho_n^t(\lambda)$ and $\vartheta_n(\lambda)$.  The Bethe equations in this case then take the form
\begin{eqnarray}
\rho^t_n(\lambda)&=& \frac{n}{2\pi}-\sum_{m=1}^\infty T_{nm}*\rho_m(\lambda)
\end{eqnarray}
where for  $n>m$ the two particle scattering kernel is
\begin{eqnarray}\notag
T_{mn}(\lambda)&=&a_{n-m}(\lambda)+a_{n+m}(\lambda)+2\sum_{j=1}^{m-1} a_{n-m+2j}(\lambda),\\
T_{nn}(\lambda)&=&2\sum_{j=1}^{n}a_{2j}(\lambda),\quad T_{mn}(\lambda)=T_{nm}(\lambda).
\end{eqnarray}

We shall study the dynamics of the system quenched from the BEC state of $N$ particles given by
\begin{eqnarray}\label{eq:Psi0LL}
\ket{\Psi_{0,N}}&=&\frac{{b^\dag_0}^N}{\sqrt{N!}}\ket{0},\\
b_0^\dag&=&\frac{1}{\sqrt{L}}\int_0^L {\rm d}x\,b^\dag(x)
\end{eqnarray}
which is the ground state of the model at $c=0$ and is also an eigenstate of $\hat{N}$.   When considering just the subsystem however $\rho_A(0)$ contains states in all particle number sectors less than $N$ and the charge  probability distribution can be determined by means of a simple argument.  Since the wave function is constant the probability of measuring a charge in $A$ for the system with $N=1$ is given by $\ell/L$.  For higher particle number since the state has no spatial correlations the detection of a boson is an  independent event with a constant rate $\ell N/L$ and therefore has a 
binomial distribution.  In the thermodynamic limit  we take $N,L\to \infty$ with $d=N/L$ the density held fixed and thus we end up with a Poisson distribution with rate $\ell d$ and $\left<N^m_A\right>^c_{0}=\ell d$ or $\kappa^s_m(0)=d,~\forall m$. 
 
From this observation it is possible to construct the reduced density matrix $\rho_A(0)$.   Introducing the boson on the restricted space $\bar{b}^\dag_0=\frac{1}{\sqrt{\ell}}\int_{x\in A}{\rm d}x \,b^\dag(x)$ we have that
\begin{eqnarray} 
\rho_A(0)=e^{-\ell d}\sum_{n=0}^\infty \frac{(\ell d)^{2n}}{n!^2}{{\bar{b}}^\dag_0}{}^n\ketbra{0_A}{0_A\!}{{\bar{b}_0}}{}^n
\end{eqnarray} 
where $\ket{0_A}$ is the vacuum inside $A$. This has the required Poisson distribution of charge and also has the same correlation within the subsystem as $\ketbra{\Psi_0}{\Psi_0}$. 

The entanglement entropy between $A$ and $\bar A$ can also be determined from this expression and originates solely from the charge fluctuations in the subsystem. The von Neumann entanglement entropy is given by Shannon entropy of the charge probability distribution and so scales as $S_A\simeq \frac{1}{2}+\frac{1}{2}\log 2\pi \ell d$.

\subsection{Gaudin-Yang model}
\label{subsec:GY}
For the case of interacting spinful fermions in one dimension  we shall study  the Gaudin-Yang  model which is also widely used to describe cold atom gas experiments. The Hamiltonian is 
\begin{equation}\label{eq:GYHam}
H_{GY}\!=\!\!\int_0^L \! \!\!{\rm d}x\, \psi_\sigma^\dag(x)\!\!\left[\!-\frac{\partial_x^2}{2m}\right]\!\!\psi_\sigma(x)+c \psi_\uparrow^\dag(x)\psi_\uparrow(x)\psi_\downarrow^\dag(x)\psi_\downarrow(x)
\end{equation}
where $\psi_\sigma^\dag(x),\psi_\sigma(x)$ are two species of fermionic operators with $\sigma=\uparrow,\downarrow$ which obey $\{\psi^\dag(x)_\sigma,\psi_{\sigma'}(y)\}=\delta_{\sigma\sigma'}\delta(x-y)$.  These describe spin $1/2$ fermions of mass $m$ interacting via a local inter-species density-density interaction of strength $c$.  As before we shall take $m=1/2$ and assume periodic boundary conditions but we shall consider both repulsive $c>0$ and attractive $c<0$ interactions.  In this case there are two $U(1)$ conservation laws corresponding to particle number  $\hat{N}=\int{\rm}d x \,\psi_\uparrow^\dag(x)\psi_{\uparrow}(x)+\psi_\downarrow^\dag(x)\psi_{\downarrow}(x)$ and magnetization $\hat{M}=\int{\rm}d x \,\psi_\uparrow^\dag(x)\psi_{\uparrow}(x)-\psi_\downarrow^\dag(x)\psi_{\downarrow}(x)$, however we shall only concentrate on the former.  

The quasi-particle content of the model depends on whether one is in the repulsive regime or attractive.  In the repulsive regime there are an infinite number of species and we denote their distributions by $\rho(\lambda),\rho^h(\lambda),\rho^t(\lambda),\vartheta(\lambda)$ and $\sigma_n(\lambda),\sigma^h_n(\lambda), \sigma^t_n(\lambda),\vartheta_n(\lambda)$ where $n\in\mathbb{N}$.  The former types of quasi-particles are associated to spin up fermions, they have unit  charge and magnetization and momentum and energy $p(\lambda)=\lambda,\epsilon_n(\lambda)=\lambda^2$.  The latter quasi-particle types, also called strings are associated to the spin degrees of freedom, they carry zero charge, magnetization $-2n$ and have zero energy and momentum. The integral equations for these distributions are 
\begin{eqnarray}
\rho^t(\lambda)&=&\frac{1}{2\pi}+\sum_{n=1}^\infty a_n*\sigma_n(\lambda)\,,\\
\sigma^t_n(\lambda)&=& a_n*\rho(\lambda)-\sum_{m=1}^\infty T_{nm}*\sigma_m(\lambda)\,,
\end{eqnarray}
where for  $T_{nm}(\lambda)$ was introduced above
In terms of these distributions the $U(1)$ charges of a state are given by $N=\int {\rm d}\lambda \rho(\lambda)$  and $M=N-2\sum_{n=1}^\infty n\int {\rm d}\lambda\,\sigma_n(\lambda)$.   

In the attractive regime there is an additional type of quasi-particle which is a bound state of two fermions forming a spin singlet. We denote its distributions by $\tilde{\rho}(\lambda)\tilde{\rho}^h(\lambda),\tilde{\rho}^t(\lambda),\tilde{\vartheta}(\lambda)$. They carry  charge $2$ and magnetization $0$ while their momentum and energy are $\tilde{p}(\lambda)=2\lambda,~\tilde{\epsilon}(\lambda)=2\lambda^2$. The corresponding integral equations are 
\begin{eqnarray}
\rho^t(k)&=&\frac{1}{2\pi}- a_1*\tilde{\rho}(\lambda)-\sum_{n=1}^\infty\, a_n*\sigma_n(\lambda)\\
\tilde{\rho}^t(\lambda)&=&\frac{1}{\pi}-a_2*\tilde{\rho}(\lambda)- a_1*\rho(\lambda)\\
\sigma^t_n(\lambda)&=&a_n*\rho(\lambda)-\sum_{m=1}^\infty T_{nm}*\sigma_m(\lambda).
\end{eqnarray}

For the fermionic model we shall again study the dynamics emerging from a BEC state wherein the bosons are formed from pairs of fermions in a singlet at the same point in space,
\begin{eqnarray}\label{eq:Psi0GY}
\ket{\Phi_{0,N}}&=&\frac{{c^\dag_0}^N}{\sqrt{N!}}\ket{0},\\
c_0^\dag&=&\frac{1}{\sqrt{\mathcal{N}_L}}\int_0^L{\rm d}x\, \psi_\uparrow^\dag(x)\psi_\downarrow^\dag(x)
\end{eqnarray}
 where $\mathcal{N}_L$ is a normalization factor. In contrast to the previous case this is not an eigenstate of the model at any $c$ but is an eigenstate of particle number $\matrixel{\Phi_{0,N}}{\hat{N}}{\Phi_{0,N}}=2N$ and magnetization $\matrixel{\Phi_{0,N}}{\hat{M}}{\Phi_{0,N}}=0$.  Once again upon tracing out $\bar{A}$,  particle number is no longer conserved however the magnetization is still $0$.  In this instance the wavefunction is not flat due to the Pauli exclusion of the fermions however for large enough subsystem size and at finite density we can apply the same arguments as before and determine that the charge distribution is also Poisson.
In particular, it is easy to check analytically the first few cumulants, which, in the thermodynamic limit, in  fact yield $\kappa_m(0)=2 d$ if $d$ denotes the density of singlet pairs.
Furthermore  the initial reduced density matrix has the form  
\begin{eqnarray}
\rho_A(0)&=&e^{-\ell d}\sum_{n=0}^\infty \frac{{(\ell d)}^{2n}}{n!^2}{{\bar{c}}^\dag_0}{}^n\ketbra{0_A}{0_A}{{\bar{c}}_0}{}^n,\\
\bar{c}^\dag_0&=&\frac{1}{\sqrt{\mathcal{N}_\ell}}\int_{x\in A}\!\!\!\!{\rm d}x\,\psi_\uparrow^\dag(x)\psi_\downarrow^\dag(x) \,\,.
\end{eqnarray} 
In other words, we have again obtained constant cumulants, which can be attributed to a Poisson distribution. It is, nevertheless, important to recall the fact that for any subsystem, the total magnetization is zero, which means that the distribution is only Poissonian for the effective `quasi-bosons'. Therefore the fermionic distribution is microscopically different, namely it has vanishing probabilities for odd fermion numbers. In the thermodynamic limit, this distribution can have the same cumulants and scales to the same limiting continuous PD, determined by the rate function of the Poisson distribution with density $2d$. We shall revisit this feature in Sec. \ref{PDs}.  The von Neumann entanglement entropy can be obtained here also and similarly coincides with the Shannon entropy of the Poisson distribution. 

\begin{figure}[t]
\centering
\includegraphics[trim= 700 0 0 0,angle=270,origin=c,width=\columnwidth]{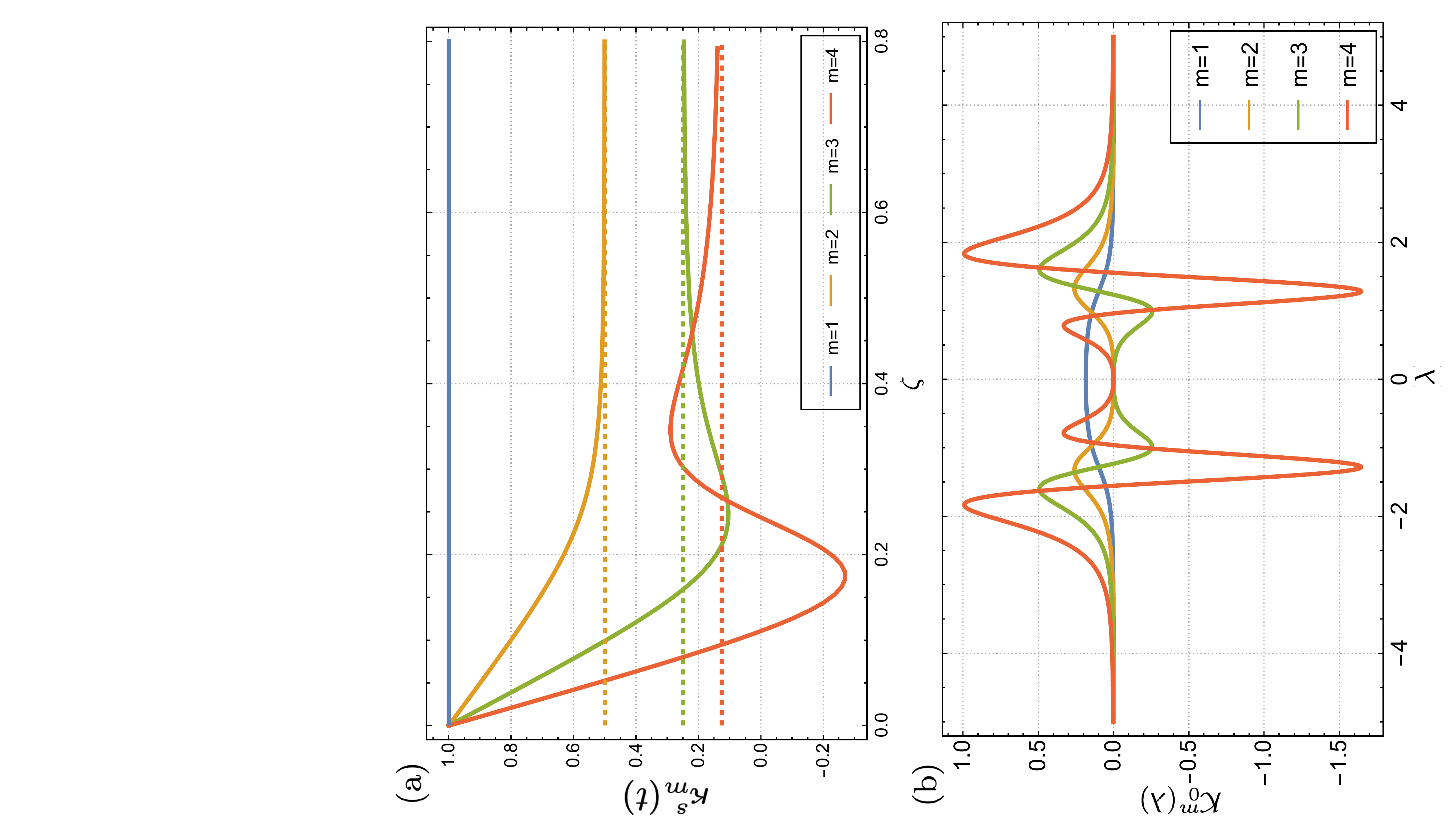}
\caption{\label{fig:LL1}(a) The scaled cumulants, $\kappa_m^s(t), ~m=1,\dots 4$ corresponding to curves with decreasing values at the $\zeta=0.2$ intersection as a function of rescaled time $\zeta=t/\ell$ for $d=1$ and $c=1$. The dashed lines are the asymptotic GGE values. 
(b) The rapidity resolved scaled cumulants, $\mathcal{K}^m_{0}(\lambda)$ as a function of $\lambda$ for $m=1,\dots,4$ corresponding to curves with increasing values at the $\lambda=-2$ intersection for $c=1,d=1$.}
\end{figure}
\begin{figure}[t]
\centering
\includegraphics[trim= 700 0 0 0,angle=270,origin=c,width=\columnwidth]{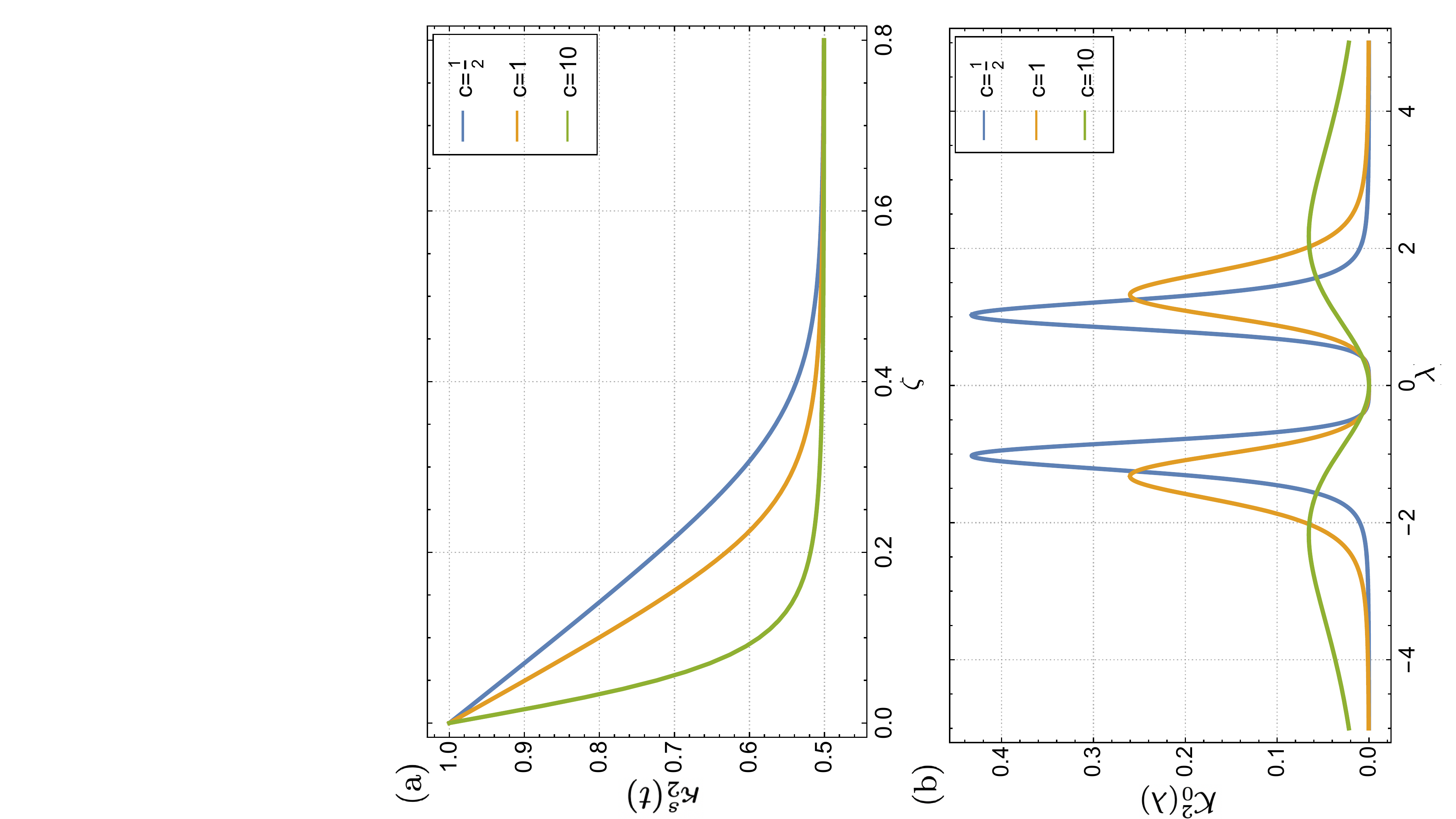}
\caption{\label{fig:LL2} (a) The scaled second cumulant, $\kappa^s_2(t)$  as a function of rescaled time $\zeta=t/\ell$ for $d=1$ and $c=\frac{1}{2},1,10$ corresponding to curves with decreasing values at the $\zeta=0.2$ intersection. (b) The rapidity resolved cumulant, $\mathcal{K}^2_{0}(\lambda)$ as a function  of $\lambda$ for $c=\frac{1}{2},1,10$ corresponding to curves with decreasing values at the $\lambda=-1$ intersection.}
\end{figure}
\begin{figure}[t]
\centering
\includegraphics[trim= 700 0 0 0,angle=270,origin=c,width=\columnwidth]{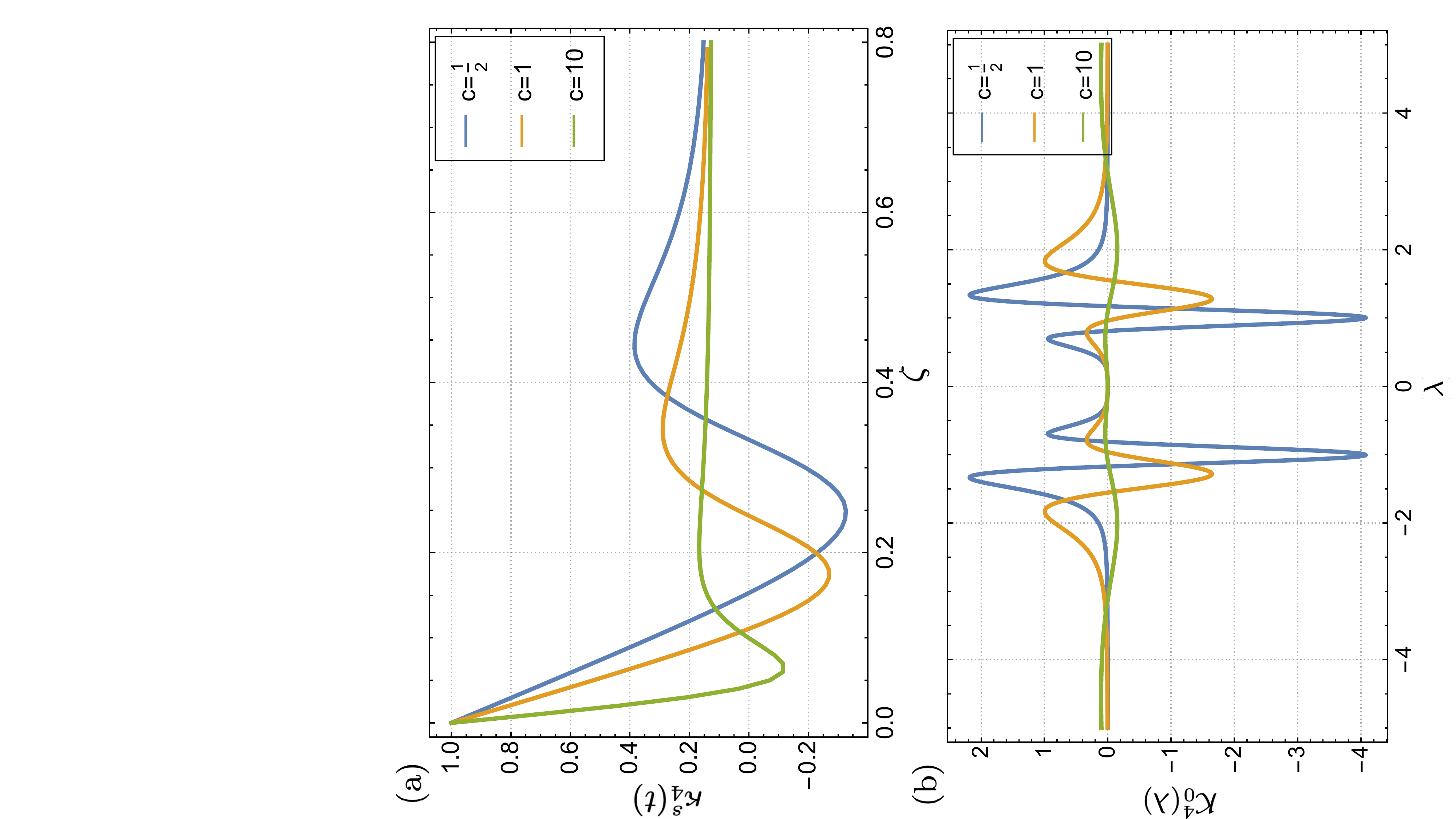}
\caption{\label{fig:LL3} (a) The  fourth scaled cumulant, $\kappa^s_4(t)$ as a function of rescaled time $\zeta=t/\ell$ for $d=1$ and $c=\frac{1}{2},1,10$ corresponding to curves with decreasing values at the $\zeta=0.4$ intersection. (b) The rapidity resolved cumulant, $\mathcal{K}^4_{0}(\lambda)$ as a function  of $\lambda$ for $c=\frac{1}{2},1,10$ corresponding to curves with increasing values at the $\lambda=-1$ intersection.}
\end{figure}

\section{FCS in the Lieb-Liniger model}
\label{sec:FCSLL}
We now turn to the explicit calculation of the SCGF in the  Lieb-Liniger model treating the repulsive and attractive regimes separately. The quench dynamics emerging from the state~\eqref{eq:Psi0LL} has been studied previously in both the repulsive~\cite{denardis2014solution} and attractive cases~\cite{piroli2016mulitparticle,piroli2016quantum}
with many properties already determined analytically and with thorough discussion on the 
 steady-state GGE in the repulsive case \cite{Palmai-Konik}. Following,  we shall make use of these exact results to obtain the FCS. 

\subsection{Repulsive interactions}\label{sec:RepLL}
The one remaining ingredient that is required to determine the FCS is the overlap function~\eqref{eq:overlapgeneral}.  
For the particular initial state~\eqref{eq:Psi0LL} this has been obtained in~\cite{brockmann2014overlaps}. In the repulsive case where there is only a single quasi-particle species this is given by
\begin{eqnarray}\label{eq:LLoverlapfun}
g(\lambda)=\log\left[\frac{\lambda^2(\lambda^2+(c/2)^2)}{d^2c^2}\right].
\end{eqnarray}
As a result the function
$C^s_\text{GGE}(\beta)$ is given by 
\begin{align}\label{eq:LLfun}
C^s_\text{GGE}(\beta)=\!\int\!\! {\rm d}\lambda\rho^t \!\!\left[\log\left(\frac{1+\eta^{-1}_{\beta}}{{1+\eta^{-1}_{0}}}\right)+\vartheta \log\frac{\eta_{\beta}e^{\beta}}{\eta_{0}}\right]
\end{align}
where $\eta_{\beta}$ is given by the solution to
\begin{align}
\log \eta_{\beta}=\log\left[\frac{\lambda^2(\lambda^2+(c/2)^2}{d^2c^2e^{\beta}}\right] - a_2*\log[1+\eta^{-1}_{\beta}(\lambda)]
\end{align}
and as stated above $\vartheta=1/(1+\eta_0)$. This type of integral equation admits an analytic solution~\cite{denardis2014solution} which is given by
\begin{align}
\eta_{\beta}(\lambda)=\frac{e^{-\beta/2}\lambda\sinh(2\pi\lambda/c)}{2\pi d I_{1-2i\lambda/c}\left(4\sqrt{\frac{de^{\beta/2}}{c}}\right)I_{1+2i\lambda/c}\left(4\sqrt{\frac{de^{\beta/2}}{c}}\right) }\,,
\label{LLExactEta}
\end{align}
where $I_m(x)$ is the modified Bessel function of the 1st kind. From this we can determine exactly both the FCS in the initial state and in the GGE.  It can then be checked that in the initial state we recover  $\left<N^m_A\right>^c_{0}=\ell d +o(\ell)$. This statement holds true for arbitrary interactions $c>0$ as can be confirmed numerically, but the result 
can be particularly easily seen in the Tonks-Girardeau limit of $c\to\infty$ in which case $\eta_\beta=e^{-\beta}{\lambda^2}/{4 d^2}$.  Accordingly we have that for the initial state in this limit
\begin{align}\label{eq:TGf}
C^s_0(\beta)|_{c\to \infty}=\int \frac{{\rm d}\lambda }{4\pi}\log\left[\frac{\lambda^2 +4d^2e^{2\beta}}{\lambda^2+4d^2}\right]=d(e^\beta-1)
\end{align}
for $\beta\in\mathbb{R}$, i.e., the SCGF of the Poisson distribution which has cumulants 
all equal.   
That is, we have recovered the know SCGF of the Poisson distribution at this limit. 
Switching now to the long time limit we have that the steady state SCGF at $c\rightarrow \infty$,
\begin{align}\label{eq:TGfSteadyState}
C^s_\text{GGE}(\beta)|_{c\to \infty}=\!\int \frac{{\rm d}\lambda }{2\pi}\log\!\left[\frac{\lambda^2 +4d^2e^{\beta}}{\lambda^2+4d^2}\right]\!=2d(e^{\beta/2}-1)\,,
\end{align}
but it is again easy to check numerically that the cumulants are in fact predicted by \eqref{eq:cumurelation} at arbitrary $c>0$. This implies that Eq. \eqref{eq:GGEsaddle} actually gives the same function $2d(e^{\beta/2}-1)$ for any $c>0$ and for $\beta\in\mathbb{R}$,  which can also be confirmed by numerical comparisons. Note that this way we confirmed the predictions of \eqref{eq:cumulantresult} and \eqref{eq:diagonalsaddle} in the repulsive regime of the LL model.

From 
these analytic results we can also write down the full time dynamics of the cumulants in the Tonks-Girardeau limit and find
\begin{eqnarray}\nonumber
\left<N^m_A(t)\right>^c & \approx & \ell d+\int\!{\rm d}\lambda\, {\rm min}[4 t|\lambda|,\ell]\left(\mathcal{K}^m_{\infty
}(\lambda)-\mathcal{K}^m_{0
}(\lambda)\right)\\\label{eq:TGfullcumu}
\mathcal{K}^m_{\infty}(\lambda)&=&\frac{\mathcal{K}^m_{0}(\lambda)}{2^{m-1}}=\partial_\beta^m\log\left[\frac{\lambda^2 +4d^2e^{\beta}}{\lambda^2+4d^2}\right]\Bigg |_{\beta=0}
\end{eqnarray}
where we used the fact that in this limit $v(\lambda)=2\lambda$. 

At finite $c$ since the function $\mathcal{K}^m_{\infty}(\lambda)$ and the quasi-particle velocity depend on the interaction strength the same relations do not hold. Instead, to analyze
the finite $c$ and $t$ behavior we plot the first four cumulants using as a function of the scaled time $\zeta=t/\ell$ in~Fig.\ref{fig:LL1} for $c=1$ and $d=10$. From this we see that while the first two cumulants are monotonic the higher cumulants are not. To understand the non-monotonicity we also plot the rapidity resolved cumulants $\mathcal{K}^m_0(\lambda)$ for $d=1,c=1$. From this we see that  $\mathcal{K}^{m\leq2}_0(\lambda)$  are positive functions while $\mathcal{K}^{m>2}_0(\lambda)$ are negative for some rapidities. By inspecting~\eqref{eq:timedepcum} and using~\eqref{eq:cumurelation} one sees that the positiveness implies that the cumulant is monotonic in time while one which is not can allow for non-monotonic behavior. In addition one can note that higher cumulants take longer to relax to their GGE values (dashed lines). This also can be understood by examining the behavior of the rapidity resolved cumulants. For $m>1$ $\mathcal{K}_0^m(\lambda)$ all vanish at the origin but have at least one set of extrema close to it. These extrema are closer to the origin at higher $m$ indicating that the slower quasi-particles contribute more to the higher cumulants resulting in a slower approach to their asymptote.  Lastly we note that for $m>1$ the number of pairs of extrema present in $\mathcal{K}_0^m(\lambda)$, $m-1$, is related to the number of extrema in $\kappa^s_m(t)$ as a function of time, namely $m-2$.

To investigate the effect of changing the interaction we plot in Fig.~\ref{fig:LL2} the second scaled cumulant $\kappa^s_2(t)$ as a function of time for $c=1/2,1,10$. From this we can also see from these plots that for lower interaction strength it takes longer to reach the asymptotic value. This could be anticipated from the fact that at $c=0$ the initial state becomes the ground state of the model. To understand this microscopically however we also plot the corresponding $\mathcal{K}^2_0(\lambda)$ as a function of $\lambda$ and see that they are peaked closer to the origin for smaller $c$. Thus the second cumulant is dominated by slower modes for lower interaction strength $c$ resulting in a slower approach to its asymptote. In Fig~\ref{fig:LL3} we plot also the fourth cumulant and see again that it approaches its asymptote slower for lower interaction strength which is a consequence of the fourth cumulant being dominated by slower modes. An analogous effect occurs if one holds $c$ constant and instead changes $d$. This is related to the anomalous relaxation phenomenon called the quantum Mpemba effect~\cite{ares2022entanglement,rylands2023microscopic,murciano2023entanglement}

\subsection{Attractive interactions}\label{sec:AttLL}
We turn now to the case of attractive interactions.  Here there are an infinite number of quasiparticle species and the overlap function for these can be obtained from~\eqref{eq:LLoverlapfun} via
\begin{eqnarray}
g_n(\lambda)=\sum_{j=1}^ng\left(\lambda+i\frac{|c|}{2}(n+1-2j)\right). 
\end{eqnarray}
The resulting cumulant generating function is 
\begin{align}\label{eq:LLattfun}
C^s_\text{GGE}(\beta)=\!\sum_n\!\int\!\! {\rm d}\lambda\rho_n^t\!\!\left[\log\!\left(\frac{1+\eta^{-1}_{n,\beta}}{{1+\eta^{-1}_{n,0}}}\right)\!+\vartheta _n\!\log\frac{\eta_{n,\beta}e^{n\beta}}{\eta_{n,0}}\right]
\end{align}
where $\eta_{n,\beta}$ are determined by
\begin{align}
\log \eta_{n,\beta}=g_n(\lambda)- \sum_{m}T_{nm}*\log[1+\eta^{-1}_{n,\beta}(\lambda)].
\end{align}
These latter equations can be rewritten using some standard TBA identities to a more convenient form~\cite{piroli2016mulitparticle,piroli2016quantum}
\begin{align}\nonumber
\log\eta_{n,\beta}=s\!*\!\log[1+\eta_{n+1,\beta}][1+\eta_{n-1,\beta}]\\
+\log\!\left[\tanh^2\left(\frac{\pi \lambda}{2 |c|}\right)\!\right]
\end{align}
where $s(\lambda)={\rm sech}\left(\frac{\pi \lambda}{2|c|}\right)$. 
 As with the repulsive case these admit an analytic solution.  It can be shown that 
\begin{align}\nonumber
\eta_{1,\beta}(\lambda)=\frac{\lambda^2/4}{|c|^2d^2 e^{\beta}(\lambda^2+|c|^2)}\!\left(\!4\lambda^4\!\!+\lambda^2(5|c|^2\!\!+\!16d e^{\beta/2}|c|)\right.\\
\left.+12|c|^2 d^2e^{\beta}+4|c|^3de^{\beta/2}+|c|^4\right)\quad
\end{align}
with the remaining functions obtained through
\begin{align}
\eta_{n,\beta}(\lambda)=\frac{\eta_{n-1,\beta}(\lambda+i|c|/2)\eta_{n-1,\beta}(\lambda-i|c|/2)}{1+\eta_{n-2,\beta}(\lambda)}-1
\end{align}
for $n>1$ and where $\eta_0(\lambda)=0$.  

These expressions become cumbersome to treat for $n>1$ 
 but simplify considerably, as does the SCGF in the infinite interaction limit $|c|\to \infty$. Taking this limit in the analytic solution one can we find that $\eta_{1,\beta}=e^{-\beta}\lambda^2/4d^2$ while all other $\eta_{n>1,\beta}$ diverge, indicating that they do not contribute to~\eqref{eq:LLattfun}. The resulting expression for the SCGF  is the same as in the Tonks-Girardeau limit~\eqref{eq:TGf} 
\begin{eqnarray}
C^s_0(\beta)|_{c\to -\infty}=C^s_0(\beta)|_{c\to \infty}
\end{eqnarray}
and so the cumulants also agree. This remains true also at arbitrary time meaning that $\kappa^s_m(t)|_{c\to \infty}=\kappa^s_m(t)|_{c\to -\infty}$.  The lack of bound-state contribution in the infinite interaction limit can be understood on energetic grounds, and the fixed energy of the initial state cannot support the formation of higher bound states. Interestingly however, in the same limit the local density-density correlation function only receives a contribution from the two-particle bound states~\cite{piroli2016mulitparticle,piroli2016quantum}. 

\begin{figure}[t]
\centering
\includegraphics[trim= 700 0 0 0,angle=270,origin=c,width=\columnwidth]{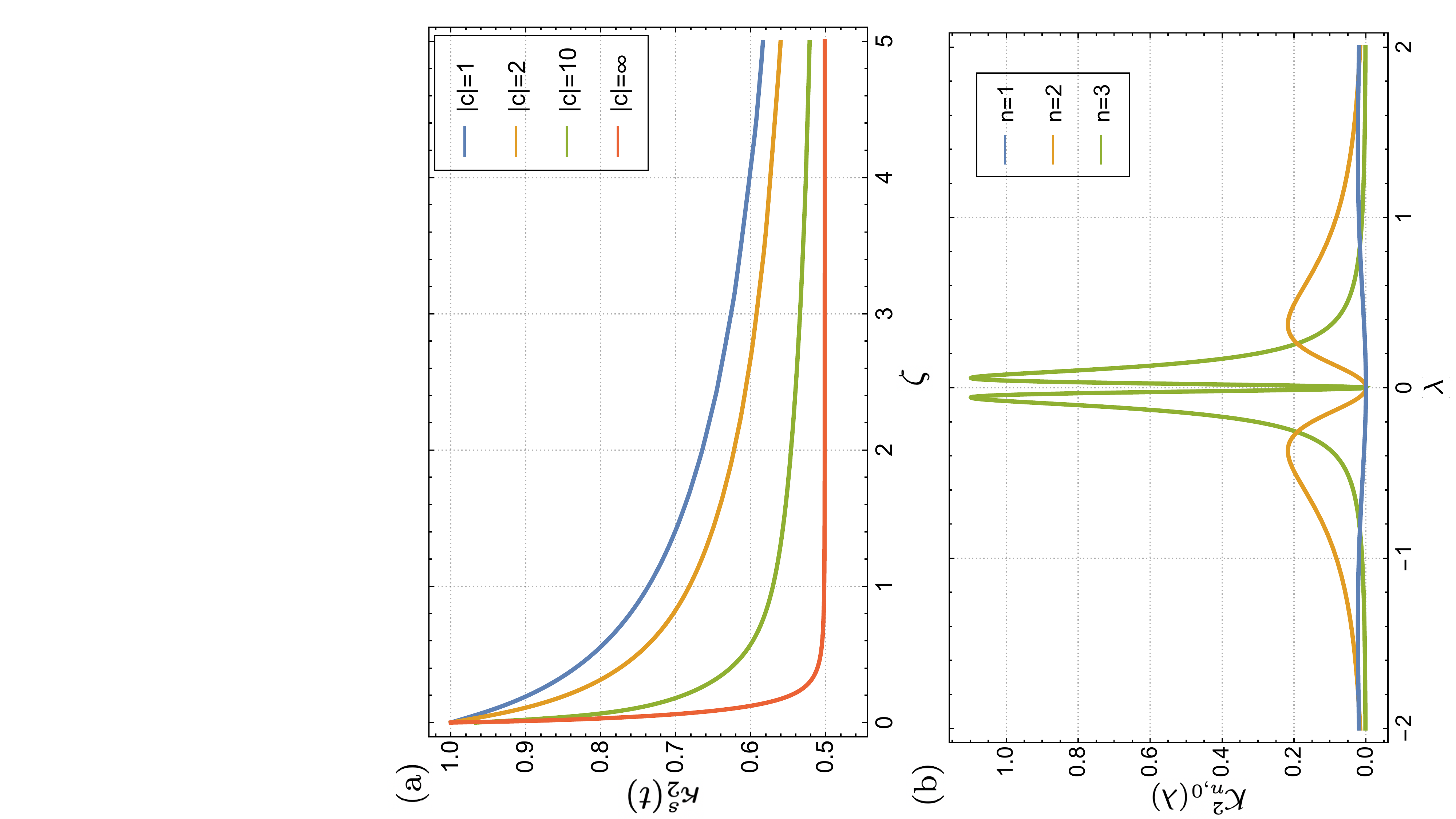}
\caption{\label{fig:attLL} (a) The scaled second cumulant, $\kappa^s_2(t)$ of the attractive Lieb-Liniger model as a function of rescaled time $\zeta=t/\ell$ for $d=1$ and $|c|=1,2,10$ corresponding to curves with decreasing values at the $\zeta=1$ intersection.  (b) The rapidity resolved cumulant, $\mathcal{K}^2_{n,0}(\lambda)$ for the first 3 quasi-particles (corresponding to curves with increasing values at the $\lambda=-0.2$ intersection) as a function  of $\lambda$ for $c=2$.}
\end{figure}

As the interaction strength is decreased bound states of increasing length contribute to the SCGF. In Fig.~\ref{fig:attLL} we plot the second cumulant $\kappa_2^s(t)$ as a function of time for different interaction strengths. We see that in contrast to the repulsive case the attractive system takes much longer to relax to its asymptotic value.   This is a result of the contribution of the bound states to the cumulants being dominated by the slowest quasi-particles. This is seen also in Fig~\ref{fig:attLL} where we plot $\mathcal{K}^2_{n,0}(\lambda)$ for $n=1,2,3$ at $|c|=2$ and see that as $n$ increases the mode resolved cumulant becomes more peaked about $\lambda=0$. Similar behavior is also seen for the higher cumulants.

\section{FCS in the Gaudin-Yang model}
\label{sec:FCSGY}
We now consider the FCS in the Gaudin-Yang model,  again splitting our analysis into two parts dealing with the repulsive and attractive regimes separately.  In both cases the key component of our analysis, the overlap functions have been obtained previously by considering appropriate limits of the integrable quenches in the Hubbard model~\cite{rylands2022integrable,rylands2023solution}.

\subsection{Attractive interactions}\label{subsec:attGY}
We start by considering the attractive regime and taking into account the quasi-particle content of the model reviewed in Sec.~\ref{subsec:GY}  we find
\begin{align}\nonumber
C^s_\text{GGE}(\beta)=\!\int\!\! {\rm d}\lambda\,\rho^t \!\left[\log\!\left(\frac{1+\zeta^{-1}_{\beta}}{{1+\zeta^{-1}_{0}}}\right)\!+\vartheta\log\frac{\zeta_{\beta}e^{\beta}}{\zeta_{0}}\right]\quad\quad\quad\quad\\\nonumber
+\!\sum_{n=1}^{\infty}\!\int {\rm d}\lambda\,\sigma^t_n \!\left[\log\left(\frac{1+\eta^{-1}_{n,\beta}}{{1+\eta^{-1}_{n,0}}}\right)+\vartheta_n \log\frac{\eta_{n,\beta}}{\eta_{n,0}}\right]\\
+\int\!\! {\rm d}\lambda\,\tilde\rho^t \!\left[\log\left(\frac{1+\tilde\zeta^{-1}_{\beta}}{{1+\tilde\zeta^{-1}_{0}}}\right)+\vartheta\log\frac{\tilde\zeta_{\beta}e^{2\beta}}{\tilde\zeta_{0}}\right],\quad\quad
\label{FCSGGEGYAtt}
\end{align}
where the functions $\eta_{n,\beta},\zeta_{\beta},\tilde{\zeta}_\beta$ satisfy the integral equations
\begin{eqnarray}\nonumber
\log\eta_{n,\beta}&=&\log\!\left[\tanh^2\left(\frac{\pi \lambda}{2 c}\right)\!\right]\!-\delta_{n,1}s\!*\!\log[1+\zeta^{-1}_\beta]\\\label{eq:GYattOTBA1}
&&+s\!*\!\log[1+\eta_{n+1,\beta}][1+\eta_{n-1,\beta}]\\\label{eq:GYattOTBA2}
\log \zeta_{\beta}&=&\log\!\left[\coth^2\left(\frac{\pi \lambda}{2 c}\right)\!\right]\!+s\!*\!\log\left[\frac{1+\tilde\zeta_\beta}{1+\eta_{1,\beta}}\right]\\\nonumber
\log\tilde{\zeta}_{\beta}&=&\log\left[\frac{\lambda^4(\lambda^2+c^2)}{(2d)^4e^{2\beta}(\lambda^2+(c/2)^2)}\right]+a_1\!*\!\log[1+\zeta^{-1}_\beta]\\\label{eq:GYattOTBA3}
&&+a_2\!*\!\log[1+\tilde{\zeta}_{\beta}^{-1}].
\end{eqnarray}
Unlike in the LL model  these equations do not admit an analytic solution that we are aware of and so must be treated numerically. There are however two limiting cases of interest.
The first is the non-interacting limit $c\to 0^-$ in which case the dependence on the $\eta_{n,\beta}$ functions drop out 
 and we find~\cite{takahashi1999thermodynamics} $\zeta_\beta^{-1}=(1+\tilde{\zeta}^{-1}_\beta)^2-1=e^{\beta}(2d)^2/\lambda^2$. As a result 
\begin{align}\label{eq:0plusf}
C^s_0(\beta)|_{c\to 0^-}=\int \frac{{\rm d}\lambda }{2\pi}\log\left[\frac{\lambda^2 +4d^2e^{2\beta}}{\lambda^2+4d^2}\right]=2d(e^\beta-1)\,,
\end{align}
for $\beta\in \mathbb{R}$, which is the SCGF of the Possion distribution with rate $2d$ and is therefore twice the result we found in the Tonks-Girardeau limit of the LL model~\eqref{eq:TGf}.  Accordingly we immediately find that the cumulants are  $\kappa^s_m(0)=2 d$ where we recall that the total fermionic density is $2d$.  Additionally, the model has a
second analytically tractable limit of interest,  $|c|\to-\infty$.  In this case not only do $\eta_{n,\beta }$ not contribute but now neither does $\zeta_\beta$ and the FCS are determined solely by the two-fermion bound states.  This results from the fact that in the strongly attractive limit the fermion pairs in the initial state which make up $c^\dag_0$ form bound states which cannot be broken apart. Thus the long time steady state consists only of these quasi-particles and no others. For this we find $\tilde{\zeta}_\beta=e^{-2\beta}\lambda^4/4d^4$ and the corresponding initial state FCS are 
\begin{align}\label{eq:Infminusf}
C^s_0(\beta)|_{c\to-\infty }=\int \frac{{\rm d}\lambda }{2\pi}\log\left[\frac{\lambda^4 +4d^4e^{4\beta}}{\lambda^4+4d^4}\right] =2d(e^\beta-1)\,, 
\end{align}
for $\beta\in\mathbb{R}$, i.e., the same result as in the free fermion limit, namely the SCGF of the Poisson distribution. 

In these two limits the SCGF can similarly be computed at the steady-states, namely we have
\begin{align}\label{eq:0plusfSteadyState}
C^s_\text{GGE}(\beta)|_{c\to 0^-}=\int \!\frac{{\rm d}\lambda }{\pi}\log\!\left[\frac{\lambda^2 +4d^2e^{\beta}}{\lambda^2+4d^2}\right]\!=4d(e^{\beta/2}-1)\,,
\end{align}
and 
\begin{align}\label{eq:InfminusfSteadyState}
C^s_\text{GGE}(\beta)|_{c\to-\infty }=\!\int \!\frac{{\rm d}\lambda }{\pi}\log\!\left[\frac{\lambda^4 +4d^4e^{2\beta}}{\lambda^4+4d^4}\right] \!\!=\!4d(e^{\beta/2}\!-1)\,, 
\end{align}
for $\beta\in \mathbb{R}$, which equal the SCGF of the LL model at infinite repulsion upon the substitution $d \rightarrow 2d $.

To investigate the cumulants at finite interaction we numerically integrate Eqs.~\eqref{eq:GYattOTBA1}-\eqref{eq:GYattOTBA3} which requires that we truncate the number of strings which are included i.e. $\eta_{n,\beta}(\lambda),\sigma_n(\lambda)=0,\forall n>N_{\rm string}$ and also that we impose a cutoff on the allowed rapidities, $|\lambda|\leq \Lambda$. 
Doing so we can confirm that the cumulants of the steady state in fact do not depend on the interaction strength. Similarly to the case of the LL model, this implies that \eqref{FCSGGEGYAtt} yields $4d(e^{\beta/2}-1)$ with $d$ denoting the density of fermions, which can be confirmed numerically.

\begin{figure}[t]
\centering
\includegraphics[trim= 700 0 0 0,angle=270,origin=c,width=\columnwidth]{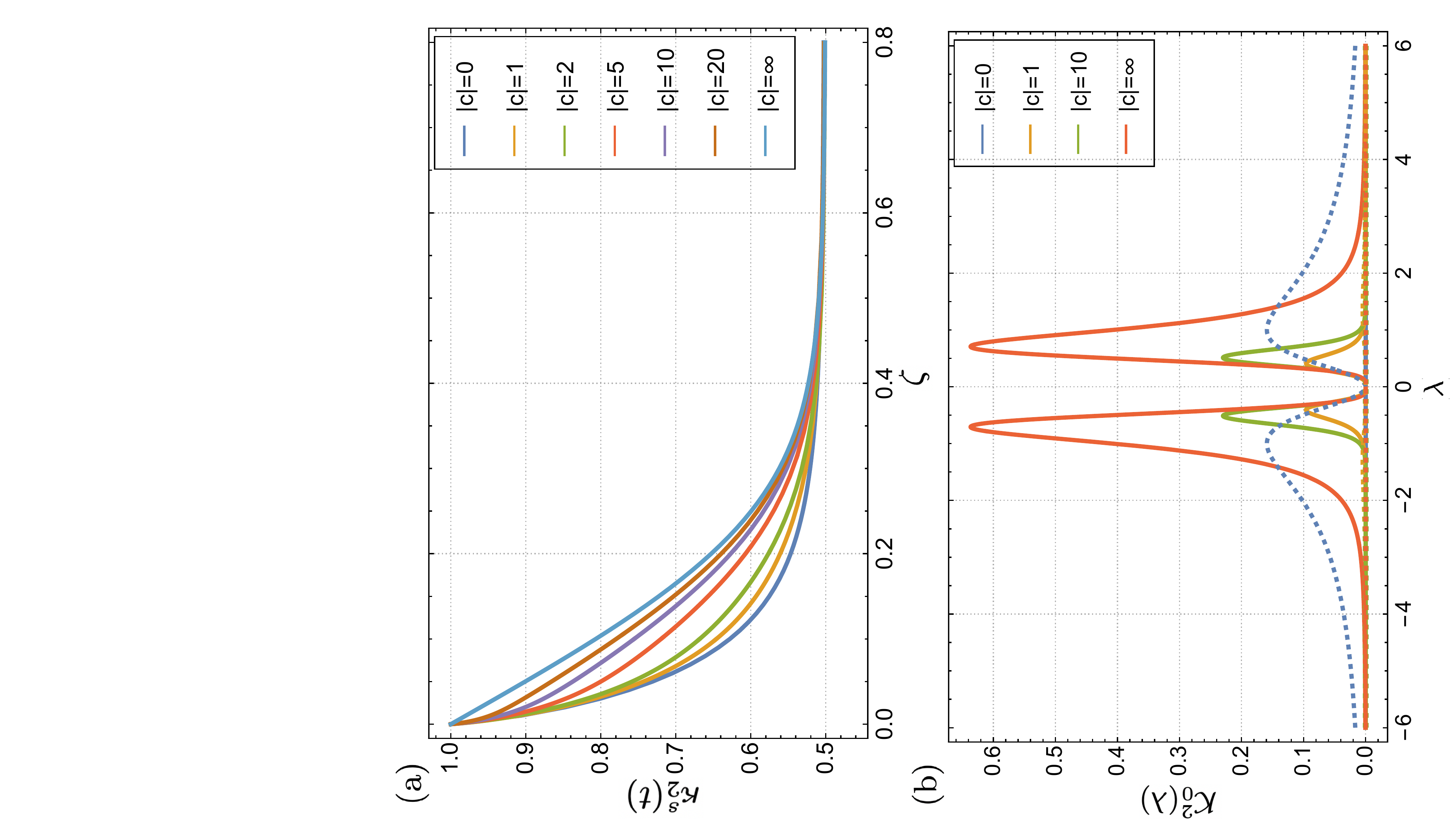}
\caption{\label{fig:AttYG1} (a) The scaled second cumulant, $\kappa^s_2(t)$ as a function of the scaled time $\zeta=t/\ell$ for different values of the interactions strengths in the attractive regime. The increasing interaction strengths correspond to curves with increasing values at the $\zeta=0.2$ intersection. (b) The rapidity resolved second cumulant for the two particle bound states (solid lines, increasing interaction values corresponding to higher peaks), $\tilde{\mathcal{K}}^2_0(\lambda)$ as a function of $\lambda$ for different interaction strengths. Also shown is the rapidity resolved second cumulant for the unbound states (dashed lines),  $\mathcal{K}^2_0(\lambda)$. In the $c=0^-$ case only unbound states are present (sole dashed curve distinguishable from the $\lambda$ axis) while in the opposite limit $|c|\to \infty$ only bound states are present. For any finite interaction the unbound states are strongly suppressed and not discernible on the same scale. }
\end{figure}

Finally, the full time evolution of the cumulants in the non-interacting limit can then be found using the previously result of~\eqref{eq:TGfullcumu} and similarly for $c\to-\infty$ whereas at finite $c$ we again resort to numerical integration of the TBA system. 
In Fig.~\ref{fig:AttYG1} we plot the second cumulant as a function of time for different interaction strength. Here we note note that in contrast to the LL model the initial state is never an eigenstate of the model and so regardless of interaction strength the cumulant relaxes at finite scaled time. Nevertheless, the approach to the GGE value still depends upon the interaction strength with the different dynamics being attributed to the changing role of the quasi-particles. Specifically for large negative interaction the two particle bound states are dominant and as shown in Fig.~\ref{fig:AttYG1} their contribution is peaked nearer the slower quasi-particles leading to an almost linear initial decrease. As the interaction strength is lowered the unbound particles also start to contribute and their mode resolved cumulant is more spread in rapidity space leading to a faster initial decay. The string contribution is negligible in all cases.

\subsection{Repulsive interactions}\label{subsec:repGY}
As a final example we look at the repulsive GY model. In this regime the spectrum of the model does not include the two particle bound states and we have that
\begin{align}\nonumber
C^s_\text{GGE}(\beta)=\!\int\!\! {\rm d}\lambda\,\rho^t \!\left[\log\left(\frac{1+\zeta^{-1}_{\beta}}{{1+\zeta^{-1}_{0}}}\right)+\vartheta\log\frac{\zeta_{\beta}e^{\beta}}{\zeta_{0}}\right]\quad\quad\quad\quad\\
+\sum_{n=1}^{\infty}\int {\rm d}\lambda\,\sigma^t_n \!\left[\log\left(\frac{1+\eta^{-1}_{n,\beta}}{{1+\eta^{-1}_{n,0}}}\right)+\vartheta_n \log\frac{\eta_{n,\beta}}{\eta_{n,0}}\right]
\end{align}
where the functions $\eta_{n,\beta}$ and $\zeta_\beta$ satisfy a set of integral equations of the form~\eqref{eq:genericTBA}.  After using some standard TBA identities we find~\cite{takahashi1999thermodynamics} 
\begin{eqnarray}\nonumber
\log\eta_{n,\beta}&=&\log\!\left[\tanh^2\left(\frac{\pi \lambda}{2 c}\right)\!\right]\!-\delta_{n,1}s\!*\!\log[1+\zeta^{-1}_\beta]\\\label{eq:GYrepOTBA1}
&&+s\!*\!\log[1+\eta_{n+1,\beta}][1+\eta_{n-1,\beta}]\\\nonumber
\log \zeta_{\beta}&=&\log\!\left[\frac{\lambda^2+(c/2)^2}{\lambda^2 e^{\beta}}\right]-\mu_c+s\!*\!\log[\lambda^2(\lambda^2+(c/2)^2)]\\\label{eq:GYrepOTBA2}
&&+s\!*\!\log[1+\eta_{1,\beta}]+s\!*\!a_1\!*\!\log[1+\zeta^{-1}_\beta].
\end{eqnarray}
where $\mu_c$ is a Lagrange multiplier, introduced to fix the average density 
 to be $2d$ and which behaves as $\lim_{c\to 0} \mu_c/c^2=2d^2$. As in the attractive case  we cannot find an analytic solution to these equations however 
we can again consider the non-interacting limit $c\to 0^+$.  In this limit the dependence on the $\eta_{n,\beta}$ functions drop out and $1+\zeta^{-1}_{\beta}=[1+\frac{4d^2}{\lambda^2}e^{\beta}]^2$~\cite{takahashi1999thermodynamics}. From this we then find that
\begin{eqnarray}
C^s_0(\beta)|_{c\to 0^+}=\int \frac{{\rm d}\lambda }{2\pi}\log\left[\frac{\lambda^2 +4d^2e^{2\beta}}{\lambda^2+4d^2}\right]
\end{eqnarray}
which agrees with the expression obtained from the~\eqref{eq:0plusf}.  Thus the non-interacting limits of the FCS approached from both the repulsive and attractive sides agree.  

Unlike the attractive case the opposite limit of $c\to\infty$ presents some problems. Indeed, exactly in this limit the initial state does not appear in the Hilbert space of the Hamiltonian since it precludes the possibility of two fermions being at the same position. By considering a lattice regularization of the model and taking the limit of infinite repulsion prior to the continuum limit it can be shown that the distribution functions become constants~\cite{rylands2022integrable}. Consequently, in the repulsive GY model, as one increases the interaction strength the distribution $\rho(\lambda)$ becomes more and more spread in rapidity space. In the absence of an analytic solution to~\eqref{eq:GYrepOTBA1} and \eqref{eq:GYrepOTBA2} must be treated numerically by first imposing some cutoff on both the string number and the rapidity space as was done in the attractive regime. However, the aforementioned spreading in rapidity space of the distributions results in a strong dependence on these truncation schemes leading to unreliable data except at small $c$ where the results are qualitatively the same as the non-interacting limit. Despite this however, we expect that the relationship between the cumulants remains valid.

\section{Charge Probability
distributions}\label{PDs}

In the previous sections we demonstrated that the scaled cumulants of charge fluctuations in the steady-state are universal for BEC quenches in both the LL and GY models. That is, they do not depend on the interaction strength of these quantum gases and they are determined by scaled cumulants in the initial states, which are essentially identical in both models. 
We now address the characterization of the full charge probability distribution in the steady state of the LL model and in the initial and steady-states of the GY model. 
The standard way of achieving this task is using the G\"artner-Ellis theorem, i.e., computing the Legendre-Fenchel transform of the SCGF and using Eq. \eqref{LimitingPD}. During the previous analysis, nevertheless, we also argued that the SCGF in the steady-states of the LL and GY models can be written by the analytic function
\begin{equation}
C^s_\text{GGE}(\beta)=C^s_\text{an}(\beta)=\sum_{m=1}^\infty \frac{\kappa^s_m \beta^m}{m!}
=2d(e^{\beta/2}-1)\,,
\label{fSteadyStateUniversal}
\end{equation}
independently of the interaction  strengths, and where
the density $d$ denotes the density of bosons or the 'quasi-bosons' in the initial states of the LL and GY models, respectively.
In fact, $C^s_\text{an}$ equals Eq. \eqref{eq:TGfSteadyState}, that is, the
$C^s_\text{GGE}(\beta)|_{c\to \infty}$ expression explicitly computed from the QA equations in the LL model. Similarly, $C^s_\text{an}(\beta)$ equals  \eqref{eq:0plusfSteadyState} as well as \eqref{eq:InfminusfSteadyState} [that is, the $C^s_\text{GGE}(\beta)|_{c\to 0^-}$ and the $C^s_\text{GGE}(\beta)|_{c\to -\infty}$ limits in the GY model] apart from a factor of 2, which further underpins the validity of $C^s_\text{an}(\beta)$ and Eqs. \eqref{fSteadyStateUniversal} at arbitrary interactions. Finally we note that the numerical solutions  for the SCGF on a finite interval, originating from the QA equations also confirm \eqref{fSteadyStateUniversal}.

Given the explicit analytic expression for the SCGF of the steady-states, and focusing first on the LL model, the rate function $I(z)$ can be easily obtained, namely
\begin{equation}
I(z)=2d-2z+2z\log(z/d)\,,
\label{SSRateFunction}
\end{equation}
i.e., twice the rate function of the Poisson distribution and thus $P(\ell z, \infty)$ has non-trivial and non-Gaussian fluctuations and large deviations. The rate function above unambiguously characterizes the limiting coarse-grained PD $P(\ell z,\infty)$ in the asymptotic sense (cf. \eqref{AsymptoticEquality}).

Nevertheless, as we have anticipated, it is also instructive to approximate the steady-state PD by using
\begin{eqnarray}
P(n, \infty)\approx\int_{-\pi}^\pi\frac{{\rm d}\beta}{2\pi}e^{-i\beta n}e^{\ell C^s(i\beta)}
\label{ProbFromLogZApprox}
\end{eqnarray}
that is by neglecting the $o(\ell)$ terms in the 2nd cumulant generating function. To use \eqref{ProbFromLogZApprox} the knowledge of the 2nd SCGF is required, but given the analytic nature  of \eqref{fSteadyStateUniversal}, we can invoke that
\begin{equation}
\label{fSteadyStateUniversalComplex}
C^s_\text{an}(i \beta)=2d(e^{i \beta/2}-1)\,.
\end{equation}

\begin{figure}[t]
\centering
\includegraphics[width=\columnwidth, trim=100 0 130 0]{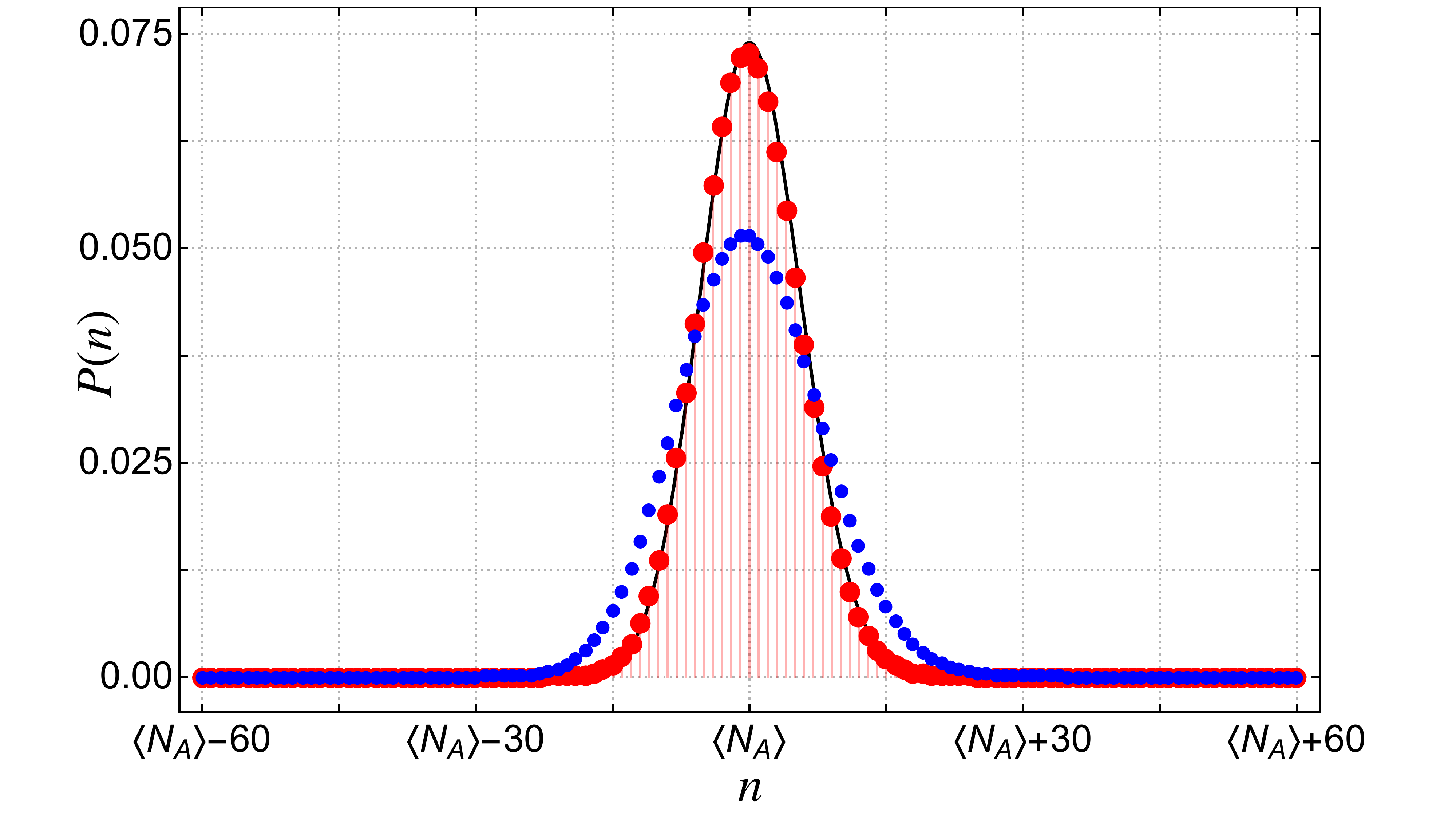}
\caption{\label{ProbA60d1LL} The probability distributions (large red  dots and black  line) of charge fluctuations in the steady state of the LL model after a BEC quench in a large but finite subsystem. The red dots were obtained using Eq. \eqref{ProbFromLogZApprox} whereas the black continuous line corresponds to the rescaled limiting PD obtained from the rate function \eqref{SSRateFunction}. The small blue dots show the Poisson PD of the initial state. The subsystem size $\ell$ equals 60 and the particle density is unity $d=1$.}
\end{figure}
In Fig. \ref{ProbA60d1LL} we display the rescaled limiting probability distribution $P(z, \infty)\asymp \text{const}\times e^{-\ell I(z/\ell)}$ obtained from the rate function and the discrete probabilities resulting from \eqref{ProbFromLogZApprox} to demonstrate their agreement.
It is important to stress, that although the PDs look Gaussian, this is not the case, since the cumulants and the asymptotics of the PDs are markedly different. One main feature that can be observed is that the variance of the PD in the steady-state is $1/\sqrt{2}$ times that of the initial state, or in other words, the distribution gets squeezed during the time evolution. 

A remark can be made in connection with approximating the steady-state PD of a large but finite system via \eqref{ProbFromLogZApprox}. Notably, the Fourier integral can be analytically computed resulting in lengthy expressions of trigonometric functions and exponential integrals. However, due to these expressions it is easy to show that $C^s_\text{an}(i\beta)$  does not define a PD in the strict sense, as for small $\ell$ negative probabilities can occur. Nevertheless increasing $\ell$, as expected, the probabilities become positive numbers sum up to one and the discrete probabilities also reproduce the predicted results for the cumulants, which we have checked explicitly focusing on the first four.

We now turn to the discussion of the PDs in the GY model and first consider the PD in the initial state. Similarly to the LL model, the scaled cumulants in the initial states are constant, that is, $\kappa^s_m(0)=2d$, where $d$ in this case is the density of fermion pairs. In the infinite subsystem size limit, the limiting continuous PD is therefore described by the rate function of the Poissonian, i.e., $I_0(z)=2d-z+z\log(z/(2d))$. Nevertheless,  the microscopic, discrete PD is anticipated to be different from a Poissonian, due the vanishing probabilities of odd particle numbers. 
As already stressed, in a strict mathematical sense our TBA/QA based methods allow only for the characterisation of the continuous, coarse-grained limiting probability distributions via the rate function. However, we can provide an approximate discrete probability distribution 
through  \eqref{ProbFromLogZApprox} that takes into account vanishing probabilities and scales to the coarse-grained PDF as well, in the following way. Namely, when directly computing the 2nd SCGF at the free fermion limit using explicitly the integral in \eqref{eq:0plusf} at imaginary $\beta$, 
the function denoted by $C^s_{0,GY}$ turns out to have the following properties.
It is a non-analytic function, equals $2d(e^{i\beta}-1)$ in the $[-\pi/2,\pi/2]$ interval, and outside this interval it is defined by the relations $C^s_{0,GY}(i\pi-i\beta)=C^s_{0,GY}(i\beta)^*$ and  $C^s_{0,GY}(-i\pi+i\beta)=C^s_{0,GY}(-i\beta)^*$, which are the consequence of a particular choice for the branches of the logarithm. 
Despite the uncontrolled treatment of these branches these symmetry properties ensure that the Fourier integrals \eqref{ProbFromLogZApprox} vanish for every odd number. {In addition,} the integrals can again be explicitly evaluated in terms of analytic expressions and hence it can be checked that for large enough subsystems, the probabilities are positive, sum up to one and the scaled cumulants match the expected value $2d$. That is, real physical behavior is correctly predicted by $C^s_{0,GY}$ and \eqref{ProbFromLogZApprox}. We again stress that such a microscopic effect, i.e., the vanishing of probabilities for odd numbers, shall not affect the limiting and continuous PD $P(\ell z,0)$, however in large but finite subsystems, this feature is anticipated to be present and the corresponding discrete PD is assumed to be well approximated by $C^s_{0,GY}(i\beta)$ via \eqref{ProbFromLogZApprox}.

Finally, we discuss the charge probabilities in steady-state of the GY model after the BEC quench. 
The coarse-grained PD $P(\ell z, \infty)\asymp e^{-\ell I(z)}$ can be simply obtained by specifying that $I(z)$ equals \eqref{SSRateFunction} upon the $ d \rightarrow 2d $ substitution. In the following, we again attempt to focus on some microscopic features of the discrete charge distribution in a finite subsystem. Specifically we are interested in whether vanishing probabilities for odd particle numbers can occur in the steady-state PD as it occurred for the PD of the initial state where it was expected based on physical considerations, but was indicated by the numerically computed 2nd SCGFs as well.

In particular, the feature of vanishing odd probabilities does take place in the  $c\rightarrow -\infty$ limit, as in this case, only the fermion, spin-singlet bound states have non-vanishing densities in the steady-state. That is, the unit charge is two, hence probabilities for odd numbers must vanish 
\footnote{One subtlety in the $c \rightarrow -\infty$ limit is that the magnetic fluctuations are sub-extensive and behave exactly the same way as it was found in Ref. \cite{cecile2023squeezed}, which might lead to a breakdown of an eventual GGE description of the steady-state. However, as the charge fluctuations are not anomalous in the initial state, it is plausible to assume that the steady-state predictions for the charge are still valid.}.
\begin{figure}[t]
\centering
\includegraphics[width=\columnwidth,trim=80 0 150 0]{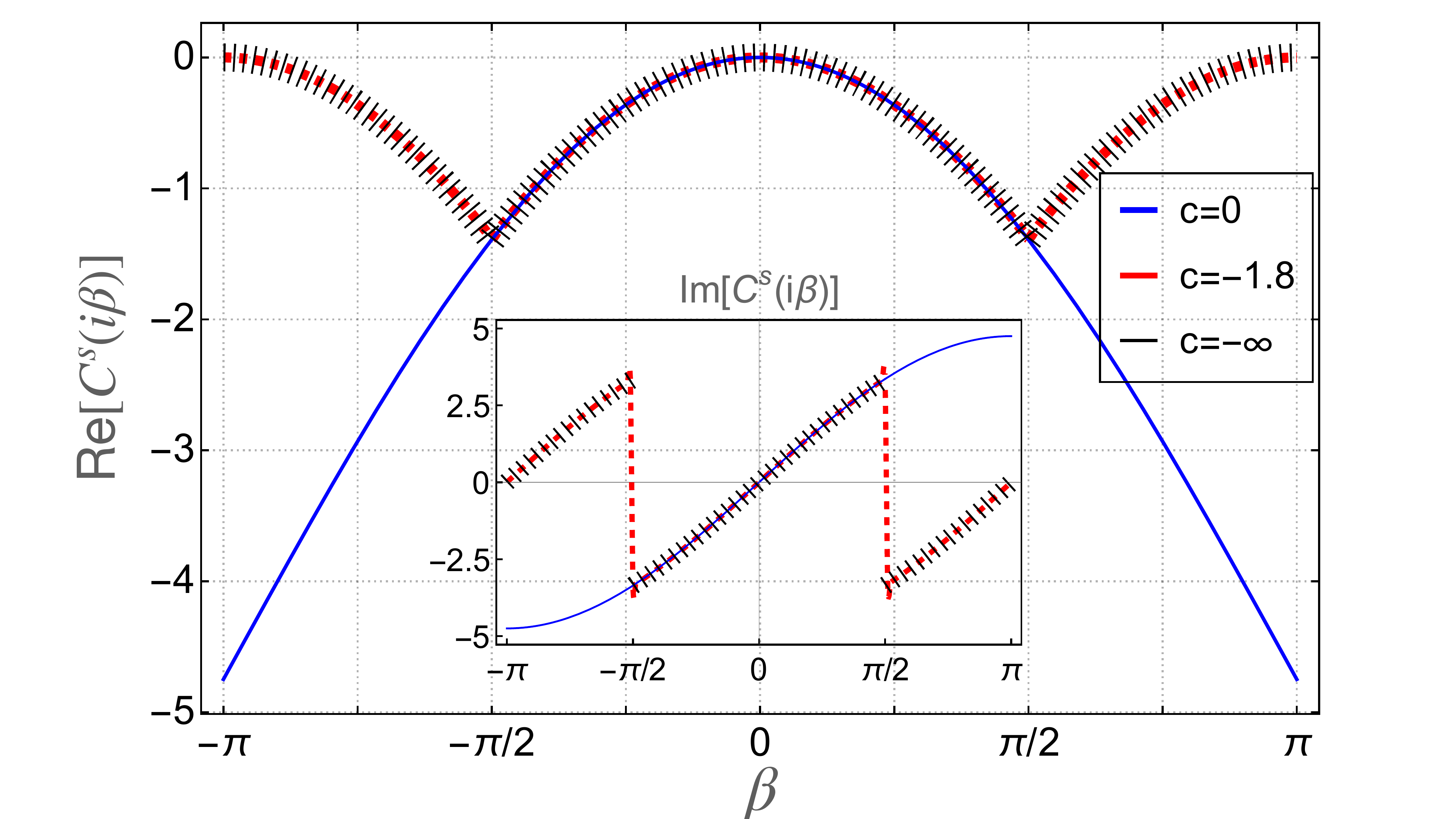}
\caption{\label{ReLogZGY} The real and imaginary (inset) parts of the 2nd scaled cumulant generating functions $C^s(i\beta)$ in the steady-state of the Gaudin-Yang model with a fixed density $d=2.4$ and various interaction strengths. The functions are obtained from the solution of the QA equations and performing numerical integration ($c=-1.8$, red dashed line) or using exact expression for the integral  \eqref{eq:GGEsaddle} ($c=0$  and $c\rightarrow -\infty$ indicated by black ticks and blue continuous line, respectively). The function for $c=0$ coincides with $C^s_\text{an}(i\beta)$, i.e., the analytic function defined by a Taylor series via the cumulants.}
\end{figure}
Indeed, when taking a look at the 2nd SCGFs originating from the QA equations and plotted in Fig. \ref{ReLogZGY}, it can be seen that in the $c\rightarrow 0^-$, i.e., the free fermion limit, the analytically computed function $C^s(i\beta)$ obtained from the integral in Eq. \eqref{eq:0plusfSteadyState} performing the $\beta\rightarrow i\beta$ substitution equals $C^s_\text{an}$. Nevertheless, for non-zero interaction strengths including the case of infinite attraction, we obtain collapsing non analytic functions. This function can be computed using the integral in \eqref{eq:InfminusfSteadyState} after substituting $\beta$ with $i\beta$. The function $C^s_{\infty,GY}$ one obtains this way equals $4d(e^{i\beta/2}-1)$ on the $[-\pi/2,\pi/2]$ interval, and outside this interval it is given by the properties $C^s_{\infty,GY}(i\pi-i\beta)=C^s_{\infty,GY}(i\beta)^*$ and  $C^s_{\infty,GY}(-i\pi+i\beta)=C^s_{\infty,GY}(-i\beta)^*$. Although the non-analytic behavior may again be attributed to the improper treatment of the branches of the logarithm, interestingly,  together with the symmetry properties of the function they capture
the aforementioned microscopic physical effect, and provide an approximate discrete PD via \eqref{ProbFromLogZApprox}. That is,  
similarly to the fermionic BEC state, the non-analytic 2nd SCGF with the particular symmetry properties imply vanishing probabilities for odd fermion numbers, when the charge fluctuations in large but finite subsystem are considered via \eqref{ProbFromLogZApprox}, which is the physically anticipated behavior at $c=-\infty$. 
Moreover, just like in the case if the free fermion BEC initial state, the Fourier integral \eqref{ProbFromLogZApprox} involving $C^s_{\infty,GY}$ can be computed analytically and when the subsystem size $\ell$ is large, the PD consists of positive probabilities summing up to one, and increasing $\ell$ the scaled cumulants converge to the prescribed value $\kappa_m^s(\infty)=2d/2^{m-1}$ as we have explicitly checked for the first four cumulants.

The situation is less understood at finite interaction strengths in the attractive regime. Although the numerically obtained 2nd SCGFs collapse to $C^s_{\infty,GY}(i\beta)$, in this case it is not clear whether or not the symmetry properties of the 2nd SCGF can be associated with a microscopic effect. Namely, it is not obvious why 
the probabilities for observing an odd number of particles in a subsystem should be zero, since at finite $c$, the root densities for the species associated with charge 1 are non-vanishing. Their contribution to the cumulants is, however, strongly suppressed for any non-zero interaction (see Fig.~\ref{fig:AttYG1}) and so it is possible that in the thermodynamic limit the odd probabilities are vanishing. Unfortunately this question cannot be definitively answered within the framework of the QA techniques we are using, therefore we leave it open for further investigations. In the repulsive regime of the model, however, we do not anticipate ambiguities and the presence of such microscopic effects due the lack of fermion bound states.

To conclude this section, we again wish to stress that by our methods we are able to give precise predictions for the continuous limiting PD $P(\ell z)$ via the rate function $I(z)$ for all the non-trivial cases, that is for the steady-state of the LL and the GY model.
Going beyond the continuous description of the limiting PD, or, eventually the $\ell\rightarrow \infty$ limit, is nevertheless out of the range of the methods we have. Therefore, we find it worthwhile to note that despite this, in many instances we could still approximate discrete and microscopic PDs describing the charge fluctuations in large but finite subsystems via \eqref{ProbFromLogZApprox}.  In particular, by enhancing the periodicity of $4d (e^{i\beta/2}-1)$ as explained above, physical microscopic structures can be recovered in PDs of the fermionic BEC initial states as well as of the steady-state of the GY model at infinite attraction which manifest in vanishing probabilities for odd charges. This periodicity property of the 2nd SCGF can emerge naturally when the function is numerically computed from the QA equations and is a result of evaluating the logarithm at complex values and hence it may be purely accidental. In fact when the derivative of the 2nd SCGF w.r.t. $\beta$ is computed at infinite attraction in the GY model, an analytic function with period $4\pi$ is obtained and hence the 2nd SCGF is assumed to inherit the same properties. Therefore,
such indications of the numerically computed 2nd SCGFs are always to be reconciled with additional physical considerations, which nevertheless are lacking at the moment for the case of the steady-state fluctuations of the GY model at finite attractive interactions.

\section{Conclusions}
\label{sec:con}
In this paper we analyzed out-of-equilibrium charge fluctuations in two paradigmatic models of one dimensional quantum gases. In particular, we investigated  the Lieb-Liniger model which describes interacting bosonic particles, and the Gaudin-Yang model in which the interacting particles are  fermions and explored their entire parameter space by considering generic repulsive and attractive interactions. We focused on two particular initial states, the Bose-Einstein condensate state and its fermionic analog for the Lieb-Liniger and Gaudin-Yang models, respectively. These choices allowed for a rather complete characterization of non-trivial charge fluctuations over the course of the entire time evolution. This achievement is due the applicability of powerful methods relying on the integrability of the physical systems and the peculiar structure of the initial states. 

In particular, using novel analytical techniques \cite{bertini2022nonequilibrium,bertini2023dynamics,Doyon_2019} and inspired by the quench action method \cite{caux2013time}, we determined all the scaled cumulants of charge fluctuations in the steady state as well their time evolution. These quantities characterize the fluctuations in a very large subsystem by keeping the leading order (linear in subsystem size) behavior. Whereas the time evolution of the scaled cumulants generically depends on the interaction strength of the models, surprisingly, their value in the steady-state was found to be independent of it. Moreover, it was also revealed that the scaled cumulants in the steady-state are uniquely characterised by the corresponding  scaled cumulants in the initial state in a remarkably simple fashion. In particular, the scaled cumulants $\kappa^s_m(0)$ in both the conventional and the fermionic BEC states are $\kappa^s_m(0)=d$ and $\kappa^s_m(0)=2d$, respectively, where $d$ is the density of bosons and fermion pairs in the two states. In the steady-states the scaled cumulants $\kappa^s_m(\infty)$ are instead given by the universal relation $\kappa^s_m(\infty)=\kappa^s_m(0)/2^{m-1}$. This relation was established based on the explicit determination of the scaled cumulants in the initial and steady-states. A formal derivation of this relationship was also provided based on comparing the full counting statistics in the diagonal ensemble and in the generalized Gibbs ensemble and by showing that the former can capture the fluctuations in the initial state.

Thanks to the exponential decay of the scaled cumulants in the steady-states we could naturally invoke the scaled cumulant generating function whose determination based on only the knowledge of the cumulants, in general, might not be straightforward. Nevertheless, the analytic function obtained this way also agrees the generating functions obtained directly from the quench action method at specific interaction strength, when analytic computations are feasible or at generic interaction strengths, when the agreement can be checked numerically. Computing the Legendre-Fenchel transform of this function we obtained an explicit 
expression for the rate function which characterizes the limiting continuous probability distribution in an infinitely large subsystem. In accordance with the scaled cumulants in the steady-state, this function predicts non-trivial and non-Gaussian fluctuations and large deviations for the charge. Additionally, we also considered the probability distributions in large but finite subsystems, where their discrete nature is still visible. An interesting feature occurs in the fermionic BEC initial state as well as in the steady-state of the Gaudin-Yang model at infinite attraction. In these cases the probabilities of observing an odd number of charge in a subsystem identically vanish. Nevertheless, resolving such discrete and microscopic features in strictly finite subsystems is beyond the scope of our current methods and hence was accomplished by relying on additional physical input.

Our work admits several pathways for further investigations. A surprising finding is the universal relationship between the scaled cumulants of the initial and steady-states, which is a consequence of the integrability of both the models and the initial states \cite{piroli2017what}. It would be important to identify the precise conditions for the onset of this phenomenon and to give a more rigorous explanation than what is presented in this work. 
A noteworthy remark is that, for initial states with vanishing scaled cumulants, such predictions clearly cannot hold.
However, universality was manifest in our examples in another way as well, namely by the lack of dependence on the interaction strength of the models. It is an interesting question whether such interaction-independence can be observed in other, possibly non-integrable models or in other quench protocols, or it again requires the integrability of both the post-quench Hamiltonian and the initial state. By applying the semi-classical arguments of the quasi-particle picture it may be possible to go beyond initial states which have a pure pair structure at least for free models as has been carried out already for entanglement dynamics~\cite{bastianello2018spreading,bastianello2020entanglement}. 

Finally we would like to highlight that similarly to the fluctuations of the charge, other conserved quantities can be investigated as well. This may require some care in the particular case of the Lieb-Liniger model, as some charges can have divergent expectation values and suitable linear combinations have to be considered \cite{Palmai-Konik}. Nevertheless, the methods applied in this work can be applied for other charges, at least for the ones with extensive initial cumulants like the energy in a subsystem.
Last but not least it  would be worthwhile to accomplish the challenging task of developing analytic or semi-analytic methods that can capture the sub-leading  behavior of fluctuations. This would be important to characterize certain microscopic effects, such as the fluctuations of conserved quantities in a subsystem for which the cumulants admit sub-extensive scaling e.g. those associated with the KPZ universality class~\cite{cecile2023squeezed}.

\begin{acknowledgments}
We wish to thank Pasquale Calabrese, Bruno Bertini, Márton Mestyán, Friedrich Hübner, Dimitrios Ampelogiannis, Giuseppe Del Vecchio Del Vecchio, Spyros Sotiriadis, Enej Ilievski, Balázs Pozsgay, Alvise Bastianello, and Benjamin Doyon for enlightening discussion on this topic and related work.  This work has been supported by Consolidator Grant No. 771536 NEMO (CR, DXH) and by the Engineering and Physical Sciences Research Council
(EPSRC) under Grant No. EP/W010194/1 (DXH).
\end{acknowledgments}
 \appendix
\section{A formal derivation of the SCGF of the initial state from the diagonal ensemble}
\label{AppA}

A main finding of this paper is the simple relationship between the scaled cumulants of the initial and the steady-states. This reads as $\kappa_m^s(\infty)=\kappa_m^s(0)/2^{m-1}$ or when promoted onto the level of the scaled cumulant generating functions, $C^s_0(\beta)=2C^s_{\text{GGE}}(\beta/2)$. These relations were obtained by comparing the FCS in the diagonal ensemble (DE) and in the GGE of the quench problem in Sec. \ref{sec:noneq} and were  justified in the following subsections by explicitly computing the corresponding cumulants in the initial and steady-states. 
While our explanation for these relationships at its present form shall remain formal, in this appendix we intend to comment more on the non-trivial bit therein, namely why and how the diagonal ensemble can describe the FCS in the initial and not in the steady-state.

First of all, let us recall the result of Ref. \cite{Doyon_2019}, which claims that in a (G)GE, the SCGF of a conserved charge can be obtained by 
\begin{equation}
C^s_\text{GGE}(\beta)=f_\text{GGE}(\underline{\beta}^{(k)}-\beta)-f_\text{GGE}(\underline{\beta}^{(k)})\,,
\end{equation}
where $f_\text{GGE}$ denotes the free energy density of a GGE characterized by the chemical potentials $\underline{\beta}^{(k)}$
and $\beta$ is associated with the particular conserved quantity of interest. We shall use this relation as a guideline in what follows.
Let us now specify the FCS in the initial state denoted by $|\Psi_0\rangle$ 
\begin{equation}
\langle\Psi_{0}|e^{\beta\hat{N}_A}|\Psi_{0}\rangle=\frac{1}{\langle\Psi_{0}|\Psi_{0}\rangle}\sum_{\Phi,\Phi'}e^{-\epsilon_{\Phi}^{*}-\epsilon_{\Phi'}}\langle\Phi|e^{\beta\hat{N}_A}|\Phi'\rangle\,,
\label{DE-FCSDerivation1}
\end{equation}
where $A$ denotes the subsystem whose length is $\ell$, $\epsilon_\Phi=-\log{\braket{\Phi}{\Psi_0}}$ are the logarithmic overlaps and in Eq. \eqref{DE-FCSDerivation1} we merely expanded the initial state in the eigenbasis of the post-quench Hamiltonian. For simplicity and transparency, we assume only one particle species. Following the logic of the  QA method, we can replace one summation by a functional integral over root distributions assuming that the size of the entire system $L$ is very large and is eventually sent to infinity. This way we may write
\begin{eqnarray}\nonumber
\langle \Psi_0|e^{\beta\hat{N}_A}|\Psi_0\rangle =\sum_{\Phi}\!\int\!\mathcal{D}[\rho]e^{S_{YY}[\rho]} \!\left[  e^{-\epsilon_{\Phi}^{*}-\epsilon[\rho]}   \langle\Phi|  e^{\beta\hat{N}_A}|\rho\rangle \right.\\
+\left.\Phi\leftrightarrow\rho\vphantom{\frac{OOOOO}{OOOOO}}\right]\times \frac{1}{\langle\Psi_{0}|\Psi_{0}\rangle},\quad\quad\quad
\label{DE-FCSDerivation2}
\end{eqnarray}
where $S_{YY}$ is the Yang-Yang entropy of the root distribution whose exponential gives the number of microstates with the same root distribution.
Note that up to this point we have two length scales, $\ell$ associated with the length of the subsystem and $L$ with the total length of the system.
The essential next step is the following: we extend the support of the subsystem $A$ and consider the charge operator in the entire system.  
This is formal step since when $L=\ell$ no charge fluctuations are expected. The step is rather based on the analogy with the treatment of FCS, more precisely the SCGF in GGEs.
Namely one can relate the SCGF of a very large subsystem ($\ell\rightarrow\infty$) with free energy densities in which the conserved quantity is regarded in the entire system. Performing this extension, it can immediately be seen that we can get rid of the second summation, since $\hat{N}$ is a conserved quantity and can have only diagonal matrix elements. That is, we can rewrite Eq. \eqref{DE-FCSDerivation2} as
\begin{equation}
\langle \Psi_0|e^{\beta\hat{N}}|\Psi_0\rangle=\frac{1}{\langle\Psi_{0}|\Psi_{0}\rangle}\int\mathcal{D}[\rho]e^{S_{YY}[\rho]-2 \text{Re}\,\epsilon[\rho]}\langle\rho|e^{\beta\hat{N}}|\rho\rangle\,,
\label{DE-FCSDerivation3}
\end{equation}
where we keep in mind that we changed the way of sending $L$ and $\ell$ to infinity by essentially equating these lengths. This expression can be further rewritten as
\begin{equation}
\begin{split}
\langle \Psi_0|e^{\beta\hat{N}}|\Psi_0\rangle=&\int\mathcal{D}[\rho]e^{S_{YY}[\rho]-2 \text{Re}\,\epsilon[\rho]+\ell\int \text{d}\lambda \beta q \rho(\lambda) }\\
&\times\left[\int\mathcal{D}[\rho]e^{S_{YY}[\rho]-2 \text{Re}\,\epsilon[\rho]}\right]^{-1}\,,
\label{DE-FCSDerivation4}
\end{split}
\end{equation}
which is equivalent to Eq. \eqref{EqSCFISPathINtegral}, if the pair-structure of the initial state is imposed. The r.h.s. of \eqref{DE-FCSDerivation4} is the rewriting of $\langle \Psi_0|e^{\beta\hat{N}}|\Psi_0\rangle$ in the diagonal ensemble. Using the saddle point approximation and taking the logarithm of \eqref{DE-FCSDerivation4}
we can write down the SCGF as a difference of two effective free energy densities, i.e., 

\begin{equation}
\begin{split}
\log\langle &\Psi_0|e^{\beta\hat{N}}|\Psi_0\rangle=\ell C^s_0(\beta)+o(\ell)
\\&=-\frac{\ell}{2}\int \text{d}\lambda \left(g(\lambda)-2 \beta q \right)\bar{\rho}_{sp}(\lambda)-\frac{1}{2}S_{YY}[\bar{\rho}_{sp}]\\
&+\frac{\ell}{2}\int \text{d}\lambda g(\lambda)\rho_{sp}(\lambda)+\frac{1}{2}S_{YY}[\rho_{sp}]+o(\ell)\,,
\label{DE-FCSDerivation5}
\end{split}
\end{equation}
where we adapted the notation of the logarithmic overlaps from Sec. \ref{sec:noneq}, took into account the $1/2$ factors due to the pair-structure of the initial state, and $\bar{\rho}_{sp}$ and $\rho_{sp}$ are the two saddle-point root distributions of the nominator and the denominator of \eqref{DE-FCSDerivation4}.

\bibliography{FCS.bib}

\end{document}